\documentclass[onecolumn]{pasj00}

\begin{document}
\SetRunningHead{Kohno et al.}{CO and NIR of NGC 5033}
\Received{2002/05/24}
\Accepted{2002/12/27}

\title{Aperture Synthesis CO($J$ = 1--0) Observations and Near-Infrared 
       Photometry of the Non-Barred Seyfert Galaxy NGC 5033}

\author{Kotaro \textsc{Kohno}}
\affil{Institute of Astronomy, The University of Tokyo, 2-21-1, Osawa, Mitaka, Tokyo 181-0015}
\email{kkohno@ioa.s.u-tokyo.ac.jp}
\author{Baltasar \textsc{Vila-Vilar\'o}}
\affil{Steward Observatory, The University of Arizona, Tucson, AZ 85721, U.S.A.}
\email{bvila@as.arizona.edu}
\author{Seiichi \textsc{Sakamoto}, Ryohei \textsc{Kawabe}, and Sumio \textsc{Ishizuki}}
\affil{National Astronomical Observatory of Japan, 2-21-1, Osawa, Mitaka, Tokyo, 181-8588}
\email{seiichi@nro.nao.ac.jp, kawabe@nro.nao.ac.jp, ishizksm@cc.nao.ac.jp}
\and
\author{Satoki \textsc{Matsushita}}
\affil{Harvard-Smithsonian Center for Astrophysics,\\ 
60 Garden St., MS 78, Cambridge, MA 02138, U.S.A.}
\email{smatsushita@cfa.harvard.edu}


\maketitle

\begin{abstract}
Aperture synthesis observations of CO($J$ = 1--0) emission
and near-infrared broad-band photometry of the non-barred Seyfert galaxy 
NGC 5033 ($D = 18.7$ Mpc) were performed.
Our \timeform{3".9} $\times$ \timeform{3".6} resolution CO 
observations reveal a perturbed distribution and the kinematics of
molecular gas in the center of NGC 5033; 
we find the characteristic gaseous features 
that are widely observed in barred spiral galaxies, 
such as two bright CO peaks near the center (separated by $\sim$ $3''$ 
or 270 pc from the nucleus), two offset ridges 
of CO emission emanating from the CO peaks, 
and a CO ring (with a radius of $\sim 14''$ or 1.3 kpc). 
Double-peaked velocity profiles are 
also evident near the two CO peaks, implying that these CO peaks 
are orbit crowding zones in a barred/oval potential.
Although NIR data only give an upper limit of the possible
bar lengths, due to a large inclination of the NGC 5033 disk ($i = 68^{\circ}$),
our CO data clearly suggests the presence of a
small (the semi-major axis of about $12''$ -- $15''$ or 1.1 -- 1.4 kpc)
nuclear bar (or oval structure) 
in the center of the ``non-barred'' galaxy NGC 5033.
Our results demonstrate that high-resolution CO imaging-spectroscopy
is useful to search for nuclear bars, even in highly inclined systems where 
isophoto fitting techniques are not applicable.
We find that the gas mass-to-dynamical mass ratio,
$M_{\rm gas}/M_{\rm dyn}$, 
is small ($\leq1$\%) within a radius of $2''$ or 180 pc,
in contrast to starburst nuclei. This implies that the
starburst does not cohabitate in the type-1.5 Seyfert nucleus of NGC 5033.

\end{abstract}

\section{Introduction}

Active galactic nuclei (AGNs) are widely believed to be powered
by accretion onto super-massive black holes (SMBHs).
Recent observational studies of galaxies have revealed that 
substantial numbers of galaxies do contain SMBHs in their nuclei 
(e.g., Kormendy, Richstone 1995; Ho 1999; \cite{vdm99}); 
SMBHs are also found even in quiescent galaxies,
such as the Milky Way (e.g., Eckart, Genzel 1996).
These facts prompt us to argue that many (or most) quiescent 
galaxies were active in the past.
It is therefore essential to investigate the fueling process
in order to understand the nature of AGNs, particularly
the difference between {\it active} and {\it starving} nuclei.

A promising fuel source is molecular gas,
because it dominates the interstellar matter (ISM) 
in the circumnuclear regions (kilo-parsec scale)
of many spiral galaxies.
It is, however, not obvious how it could be transported from
typical galactic radii ($\sim$ 10 kpc) down to the scale of the
central engine ($\leq$ 1 pc).
Galactic bars are frequently invoked as candidates for
facilitating the transfer of mass in a host of active galaxies
(e.g., Shlosman et al.\ 1989). In fact,
the bar-driven gas structures can be widely seen in many Seyfert galaxies
(e.g., \cite{hb95}; Kohno et al.\ 1999b; \cite{bak99}; \cite{mai00};
\cite{sch00}; \cite{kk01}; \cite{kod02}).
There appears, however, observational evidence {\it against} the bar-induced 
fueling scenario as well.
Ho et al.\ (1997c) have reported that galaxies hosting AGNs
do not appear to be significantly affected by the presence of a bar,
based on their extensive optical spectroscopic survey of 486 nearby galaxies.
A similar conclusion was reached (Hunt, Malkan 1999) using 891 galaxies in
an extended 12 $\mu$m sample (Rush et al.\ 1993).
We should note that their conclusion was based on an optical morphology classification
which could be severely affected by dust extinction
(e.g., \cite{sco88}; \cite{thr89}).
Near-infrared (NIR) photometric surveys have also suggested that
a significant fraction of the Seyfert galaxies show no evidence
for the presence of barred potentials
(McLeod, Rieke 1995; Mulchaey, Regan 1997),
although Seyfert host galaxies seem to be barred more often than non-active galaxies 
(\cite{ksp00}; \cite{lai02}).
High angular resolution snapshot surveys using HST NICMOS also failed 
to find any signature of
a strong bar for many of Seyfert galaxies 
(\cite{rm99}; \cite{mp99}; but see \cite{lai02}).
Given the absence of bars, then, what kind of physical processes can drive
the ISM into the nuclear region?

In order to investigate the distributions and kinematics of
the ISM in non-barred Seyfert galaxies,
we carried out high-resolution CO($J$ = 1--0) observations 
of the type-1.5 Seyfert galaxy NGC 5033 (Ho et al.\ 1997a, 1997b).
NIR broad-band photometry was also performed to determine the underlying
potential from the stellar surface density, 
and also to see the distribution of ISM via NIR color.
NGC 5033 is classified as non-barred in both the Third Reference Catalogue
of Bright Galaxies (RC3; de Vaucouleurs et al.\ 1991)
and A Revised Shapley-Ames Catalog of Bright Galaxies 
(RSA; Sandage, Tammann 1987).
The distribution and kinematics of atomic gas 
in NGC 5033 have been investigated,
and a warping of the galactic plane at larger radii ($r >$ a few kpc) is 
reported (e.g., Bosma 1981; \cite{wev86}; Thean et al.\ 1997),
but there is little evidence for a bar (Thean et al.\ 1997).
NGC 5033 is therefore an unique target to investigate the distribution 
and kinematics of ISM in non-barred Seyfert hosts.
This galaxy is also suitable for a detailed study of the structure
because of its proximity ($D$ = 18.7 Mpc; Tully 1988) and large
apparent size ($10.\hspace{-2pt}'7 \times 5.\hspace{-2pt}'0$; 
de Vaucouleurs et al. 1991).
The basic properties of NGC 5033 are listed in table 1.

We describe millimeter and NIR observations in section 2,
and present new results on CO and NIR emission in section 3. 
In section 4, we show that the molecular gas distribution and kinematics 
in the central a few kpc region of NGC 5033 are governed 
by a {\it small} nuclear bar. 
Gravitational instabilities of the molecular gas 
in the center of NGC 5033 are also discussed,
implying the absence of strong nuclear star formation
in the center of the type-1.5 Seyfert galaxy NGC 5033.
We summarize our conclusions in section 5.

\section{Observations and Data Reduction}

\subsection{CO(1--0) Data with the NMA}

The central region of NGC 5033 was observed 
in the $J$=1--0 line of CO with the Nobeyama Millimeter Array (NMA).
The observations were made during two separated runs,
from 1993 December to 1994 April for the C and D configurations,
and in 1996 January for the AB configuration.
Due to the limitation of the minimum projected baseline length (10 m),
extended structures larger than about 50$''$ in each channel map
were not sampled in the observations.
The front-ends were tunerless SIS receivers,
whose receiver temperature was about 40 K in the double side band,
and the system noise temperatures (in single side band)
were 400 -- 800 K during the observations.
A digital spectro-correlator FX was configured
to cover 320 MHz with 1024 channels per baseline.
A side-band separation was achieved by 90$^\circ$ phase switching.
We observed a continuum source, 1308+326, every 30 minutes 
to monitor the temporal variations of the instrumental complex gain. 
The passband across 1024 channels
was calibrated through observations of 3C 273.
The flux density of 1308+326 was determined 
from comparisons with planets of known brightness temperatures.
The uncertainty in the absolute flux scale is estimated to be $\sim \pm$ 20\%.
The raw data were calibrated and edited using the package UVPROC-II 
developed at NRO (\cite{tsu97}), and then Fourier transformed 
with natural weighting using the AIPS task MX.
A conventional CLEAN method was applied 
to deconvolve the synthesized beam pattern.
The CLEAN beam was \timeform{3".9} $\times$ \timeform{3".6} 
(position angle {\it P.A.} $ = 6^\circ$).
We made 19.55 km s$^{-1}$ width CO channel maps 
at an interval of 9.78 km s$^{\rm -1}$.
The typical rms noise level of each channel map is 30 mJy beam$^{\rm -1}$,
or 200 mK in brightness temperature scale, $T_{\rm b}$.
The parameters of the NMA observations are summarized
in table 2.

\subsection{$J$ and $K'$ Band Images with the OASIS}

$J$ (1.25 $\mu$m) and $K'$ band (2.15 $\mu$m) images 
of NGC 5033 were obtained 
with the 256 $\times$ 256 element HgCdTe infrared camera OASIS 
(Yamashita et al.\ 1995)
at the Cassegrain focus of the 188 cm reflector 
at the Okayama Astrophysical Observatory, 
on 1996 December 22 and 23. 
The pixel size of the camera, 40 $\mu$m, 
corresponds to $0.\hspace{-2pt}''96$ at the focal plane 
of the F/4.5 camera, providing a field of view of 
4$.\hspace{-2pt}'$1 $\times$ 4$.\hspace{-2pt}'$1.
In order to avoid saturation due to the bright Seyfert nucleus,
each frame was taken with short exposure times
(20 s for $J$ and 5 s for $K'$).
The net exposure times were 480 s for the $J$ band and 400 s
for the $K'$ band, respectively.

Data reduction was made using the NOAO IRAF package.
A dark frame was subtracted from each sky frame,
and then images were divided by a dome-flat frame. 
After the flattening, sky-subtracted object images were
aligned with respect to the stars in the field and combined. 
Bad pixels were clipped during the combining. 
For the absolute flux calibration, HD 105601,
which is listed at the UKIRT standard star catalogue, 
was observed during the same nights.
Atmospheric extinction corrections were made,
whereas no correction for Galactic
extinction was applied.
We estimate an uncertainty of a few \% in the magnitudes,
mainly due to sky subtraction errors.
The resultant spatial resolutions estimated
from the profiles of stars in the field were 
\timeform{1".7} for the $K'$ band
and \timeform{3".2} for the $J$ band images, respectively.

\section{Results}

\subsection{CO(1--0)}

\subsubsection{Channel maps}

Figure 1 shows CO channel maps from the central 
$45'' \times 75''$ region (4.1 kpc $\times$ 6.8 kpc) of NGC 5033.
We detected significant ($>$ 3$\sigma$) CO emission in 49 adjacent channels
with a velocity range of $V_{\rm LSR}$ = 651.5--1119.4 km s$^{\rm -1}$.
This velocity width (full width of zero intensity) of 479 km s$^{\rm -1}$ is
almost the same as that of the single-dish CO(1--0) profiles
(e.g., Heckman et al.\ 1989; Young et al.\ 1995; 
Braine, Combes 1993; \cite{nn01}).
In order to estimate the missing flux of our aperture synthesis observations,
we compared our data with the existing single-dish observations.
This galaxy has been observed repeatedly in the CO(1--0) line with 
various single-dish telescopes (Stark et al.\ 1987; Heckman et al.\ 1989; 
Braine et al. 1993; Young et al.\ 1995;
Elfhag et al.\ 1996; Maiolino et al.\ 1997;
Papadopoulos, Seaquist 1998; Vila-Vilar\'o et al.\ 1998; 
Curran et al.\ 2000; \cite{nn01}).
Here, we convolved our data cube to a $45''$ beam and compared the NMA flux 
with the FCRAO 14 m observations. 
The CO flux at the same position as the FCRAO observations
[$\alpha$ (B1950) =
$13^{\rm h}11^{\rm m}09^{\rm s}\hspace{-5pt}.\hspace{2pt}7$,
$\delta$ (B1950) = $+36^\circ51'30''$],
is 325 $\pm$ 8 Jy beam$^{-1}$ km s$^{-1}$, whereas the FCRAO flux is
7.2 $\pm$ 0.77 K km s$^{-1}$ in $\int T_{\rm a}^*(v)dv$ or
302 $\pm$ 32 Jy beam$^{-1}$ km s$^{-1}$ using a sensitivity factor
of 42 Jy K$^{-1}$ for the $T_{\rm a}^*$ scale. (Young et al.\ 1995).
Considering the uncertainties in the absolute flux scale ($\sim \pm 20$\%),
most of the single-dish flux seems to be recovered by our observations,
and we can safely discuss the distribution and kinematics of 
the molecular gas below.

\subsubsection{CO distribution}

A velocity-integrated intensity map of the CO emission 
is shown in figure 2b, 
with an optical image of NGC 5033 (figure 2a)
from the Digitized Sky Survey.
This CO image was obtained from the 0-th moment of the CO data cube
with the AIPS task MOMNT.
To minimize the contribution from noise, we computed a moment map
with a clip level of $2 \sigma$ in each channel map.

The most outstanding feature seen in the integrated intensity map
of CO, figure 2b, is two strong CO peaks, 
which appear to straddle the nucleus. 
These two CO peaks lie along the major axis of the galaxy ($P.A.$ = $-8^\circ$),
and are separated by $\sim 3''$ (270 pc) from the nucleus. 
We also find offset ridges of the CO emission, 
emanating from the two CO peaks,
along a $P.A.$ of about $-60^\circ$.

It is evident that there is an extended and patchy 
disk-like structure over the field of view. 
The outer boundary of the CO emission is almost elliptical,
with an axial ratio of about 0.4. 
This axial ratio can be explained as a disk
with an intrinsically circular boundary,
at an inclination angle of $\sim 66^\circ$.
This inclination angle of the molecular disk agrees with 
that of the whole galaxy (table 1; Thean et al.\ 1997).
The radius of the disk-like structure is about $30''$ = 2.7 kpc.
This size is mostly consistent with a $\sim 7''$ resolution CO map 
obtained from BIMA observations (Wong, Blitz 2002).
A careful inspection of the CO map suggests that
some of the CO patches on the ``disk'' are a part of continuous structures,  
such as a ring and a spiral arm.
The ring and spiral-arm like structures are particularly remarkable 
in the southern part of the map, as traced by the solid lines in figure 3.
These overall CO structure agree well with recent OVRO millimeter array
observations (Baker 1999).

Figure 4 shows the CO spectra from the central
$21'' \times 21''$ region of NGC 5033. 
The CO spectra near the two central peaks are characterized
by broad line widths ($\geq$ 200 km s$^{-1}$ in FWHM)
and double or multiple peaked profiles,
while the CO spectra in the disk region show single Gaussian-shaped profiles
and narrow widths ($\sim$ 50 km s$^{-1}$ in FWHM).

\subsubsection{Molecular gas mass and surface density}

The mass of the molecular gas is estimated from the CO intensity, 
using a $N_{\rm H_2}/I_{\rm CO}$ conversion factor, $X_{\rm CO}$.
Here, we adopted a Galactic $X_{\rm CO}$ value of
$1.8 \times 10^{\rm 20}$ cm$^{\rm -2}$ (K km s$^{\rm -1}$)$^{\rm -1}$
(Dame et al.\ 2001).

The surface mass density of molecular hydrogen 
on the galaxy plane is calculated as
\begin{equation}
\Sigma_{\rm H_2} = 2.89 \times \cos(i) \cdot
\left( \frac{I_{\rm CO}}{\mbox{K km s$^{-1}$}} \right)
\left[ \frac{X_{\rm CO}}{1.8 \times 10^{20}\ \mbox{cm$^{-2}$ (K km s$^{-1}$)$^{-1}$} } \right] 
\mbox{\ $M_\odot$ pc$^{-2}$},
\end{equation}
where $I_{\rm CO}$ is the CO intensity, and $i$ is the inclination of the galaxy 
($i = 68^\circ$ for NGC 5033). 
For our CO observations,
the $1 \sigma$ level of the CO integrated intensity map,
2.0 Jy beam$^{-1}$ km s$^{-1}$ or 13 K km s$^{-1}$ in $T_{\rm b}$, 
corresponds to a face-on molecular gas surface density of 
$\Sigma_{\rm gas} = 23$ $M_\odot$ pc$^{\rm -2}$
(or a H$_2$ column density of $2.4 \times 10^{21}$ cm$^{-2}$), 
and the CO peak (about $14 \sigma$) is 320 $M_\odot$ pc$^{\rm -2}$, 
including He and heavier element masses,
as $\Sigma_{\rm gas} = 1.36 \cdot \Sigma_{\rm H_2}$.
Note that the gas surface densities don't depend on 
the adopted distance to the galaxy.
The molecular gas density derived here should be compared 
with the atomic gas surface density; 
Thean et al. (1997) detected a {\it peak} H\,{\footnotesize I} column density of 
6.5 $\times$ 10$^{21}$ atoms cm$^{-2}$ 
($22.\hspace{-2pt}''7 \times 18.\hspace{-2pt}''5$ beam) 
or a {\it peak} face-on H\,{\footnotesize I} surface density, 
$\Sigma_{\rm H}$, of 18 $M_\odot$ pc$^{-2}$,
which is below the lowest contour in figure 2b and 3
(face-on H$_2$ surface density $\Sigma_{\rm H_2}$ of 25 $M_\odot$ pc$^{\rm -2}$). 
Although the $\Sigma_{\rm H}$ is an averaged value over a much larger area 
(22$.\hspace{-2pt}''$7 $\times$ 18$.\hspace{-2pt}''$5) 
than our $\Sigma_{\rm H_2}$ data (3$.\hspace{-2pt}''$9 $\times$ 3$.\hspace{-2pt}''$6),
we may conclude that molecular gas dominates the ISM 
in the central region of NGC 5033.

The molecular gas mass is then computed 
by multiplying the surface density of the gas
by area. Converting a unit of CO intensity from K km s$^{-1}$ to
Jy km s$^{-1}$ per arcsec$^{2}$, we obtain
\begin{equation}
M(\mbox{H$_2$}) = 7.0\times10^3 
\left( \frac{S_{\rm CO}}{\mbox{Jy km s$^{-1}$}} \right) 
\left( \frac{D}{\mbox{Mpc}} \right)^2
\left[ \frac{X_{\rm CO}}{1.8\times10^{20}\ \mbox{cm$^{-2}$ (K km s$^{-1}$)$^{-1}$}} \right]
\mbox{\ $M_\odot$},
\end{equation}
where $S_{\rm CO}$ is the CO flux density, and $D$ is the distance.
The observed CO flux within the central $r < 30''$ region
is 430 Jy km s$^{-1}$, which corresponds to a total molecular gas mass of
$M_{\rm gas} = 1.4 \times 10^9$ $M_\odot$, 
including He and heavier elements ($M_{\rm gas} = 1.36 \cdot M_{\rm H_2}$).

\subsubsection{Radial distribution of molecular gas}

Figure 5 shows the radial distribution of the molecular 
gas surface density in NGC 5033.
The integrated intensity of CO in figure 2b was azimuthally averaged 
over successive annuli with $1''$ = 90.7 pc width using the AIPS task IRING,
and was used to calculate the face-on molecular gas surface density 
by adopting the above formula.
The maximum gas surface density is located 
at a radius of $\sim$ 400 pc ($\sim 4.\hspace{-2pt}''4$),
not at the center. A second increase in the gas surface density is also evident 
at $r \sim 1.3$ kpc ($\sim 14''$),
which corresponds to the radius of the ``partial ring'' denoted in figure 3.

\subsubsection{Velocity field}

Figure 2c shows an intensity-weighted isovelocity contour map
of the CO emission. This map was made by computing 
the first moment of the CO data cube,
as $<\!\!v\!\!> \,\, = \Sigma_i v_i S_i/\Sigma_i S_i$.
It is evident that circular motion dominates the kinematics 
in the central a few kpc region of NGC 5033 because
a systematic change in the positions of the emission
as a function of velocity with the overall symmetry can be seen, 
suggesting an ordered circular motion. 
However, as seen in the channel maps,
this CO velocity field also reveals a ``Z-shape'' twist
of isovelocity contours near the systemic velocity, 
which is a signature of non-circular motions.
It is also clear that the isovelocity contours near the center 
tend to align with the position angle of the CO ridges ($P.A. \sim -60^\circ$), 
rather than the kinematical minor axis ($P.A. \sim -98^\circ$).
In a pure circular rotation case,
the isovelocity contours near the systemic velocity should be straight lines 
perpendicular to the kinematical major axis (i.e., aligned to the minor axis).
This deviation is also evident in the velocity channel maps 
of $V_{\rm LSR} = 866 - 895$ km s$^{-1}$ in figure 1.

We determined the kinematical parameters of the gas disk
(dynamical center, position angle of the major axis, 
inclination angle, and systemic velocity)
by a least-squares fitting of the intensity-weighted isovelocity field
as a circular rotation.
We used the GAL package in AIPS for this analysis.
The derived parameters are listed in table 3.
These results agree with values determined from the $\sim$ 20$''$ resolution 
H\,{\footnotesize I} observations (Thean et al.\ 1997).
It should be noted that the observed CO velocity field contains
local deviations from circular rotation, as described above;
however, and the kinematic parameters derived here can be affected
by the non-circular motions.
Hence,, we adopted those determined by Thean et al.\ (1997) 
in the following analysis.

\subsubsection{P--V diagram and rotation curve}

A position--velocity diagram (P--V diagram)
along the major axis ($P.A.$ = $-8^\circ$)
of NGC 5033 is displayed in figure \ref{fig:pv}. 

We determined the circular rotation curve of the inner part of NGC 5033
from our CO data by averaging the intensity-weighted mean velocities
(figure 2c) within a $\pm 5^\circ$ area along the major axis of the galaxy,
using the GAL package in AIPS.
The resultant rotation curve is shown in figure \ref{fig:rotcurve}.

The rotation velocities at the flat part are $\sim$ 220 km s$^{-1}$
on the plane of the disk (adopting $i = 68^\circ$). 
This agrees well with the $\sim$ 20$''$ resolution 
H\,{\footnotesize I} observations (Thean et al.\ 1997)
and $16''$ resolution CO observations (Sofue 1997; \cite{nis01}).

Near the center ($r < 10''$), this diagram shows
a gradual rise of the rotation velocity.
Smearing of the observing beam is a possible cause of this {\it gradual} rise,
and the true rotation velocity could be larger there.
The CO spectra near the center shows
very broad and multiple velocity components (figure \ref{fig:COspectra}),
and a simple moment analysis could fail to trace the true rotation velocities
in this case.

In fact, the P--V diagram in figure \ref{fig:pv} shows
a very steep rise of the rotation velocities near the center
if we trace the terminal velocities as the true rotation velocities;
it reaches a constant (flat rotation) at $r \sim 2''$.
Given our finite spatial resolution, the turnover radius of the rotation curve
($r \sim 2''$) could be an upper limit.

Another possibility of a {\it gradual} rise of the circular rotation velocity
is a bar; non-circular motion associated with the presence of a bar
may be the cause of a steeper P--V diagram than the actual rotation curve
(Bureau, Athanassoula 1999; Athanassoula, Bureau 1999).

It should the be noted that the velocity field contains 
significant non-circular motions,
and the derived values must be treated as an averaged rotation curve
(e.g., Sakamoto et al.\ 1999a), possibly with significant local deviations.

From these considerations, we here assume a constant rotation velocity of
220 km s$^{-1}$ for $r>2''$, 
and a rigid-body rotation of $\Omega = 1.2$ km s$^{-1}$ pc$^{-1}$
for $r<2''$. 
This rotation curve is indicated as the dashed line
in figure \ref{fig:rotcurve}, and is applied to resonance identifications
in the following discussion (subsection 4.2).

Note that Baker (1999) reported that there is a kinematically decoupled gas 
component just around the center of NGC 5033 (i.e., between the two CO peaks
in figure 2b) based on a P--V diagram of the OVRO interferometer 
CO(2--1) data; however, the CO(1--0) emission in our map 
seems to be too weak to identify such features.

\subsubsection{Velocity dispersion}

Figure \ref{fig:dispersion} shows the intensity-weighted CO velocity dispersion
along the line of sight, as derived from the second moment of the
CO data cube, $\sigma_v = \Sigma_i (v_i - <\!\!v\!\!>)^2 S_i/\Sigma_i S_i$.
It should be noted that this map contains both the intrinsic velocity
dispersion of the gas and the gradient of the rotation velocity 
in the observing beam;
the very large velocity dispersion area along the minor axis of
the galaxy represents the steep rise of the rotation curve in the central
region of the galaxy.
Near the center of the galaxy, it is evident that the second moment values 
at the two central CO peaks
are quite large ($>45$ km s$^{-1}$) compared with other regions.
The CO line profiles in this area appear to deviate from a simple
Gaussian shape, and are perhaps multiple-peaked, as can be seen in figure 4.
Therefore, the second moment values near the central CO peaks 
should not be treated as a velocity dispersion of gas, yet
complex gas kinematics near the two central CO peaks is suggested.

\subsection{NIR Images}

\subsubsection{Isophotos}

$J$ and $K'$ band images obtained with the OASIS observations 
are presented in figure \ref{fig:JandK}.
The surface brightness at the outer boundary is about 19 mag arcsec$^{\rm -1}$ 
in the $K'$ band, and 21 mag arcsec$^{\rm -1}$ in the $J$ band, respectively. 
We also show contour maps of the $K'$ band in the central $40'' \times 40''$ 
region in figure \ref{fig:Kcloseup}.
The overall $K'$ band structures agree with the previous NIR observations
(\cite{mr95}; \cite{pel99}).

In order to quantify the incidence of a bar,
we have fitted ellipses to isophotos of each band
using the routine ELLIPSE in the STSDAS package
within IRAF. 
Radial profiles of the surface brightness,
ellipticity, and position angle derived from the ellipse fits
of the $K'$ band data are given in figure \ref{fig:Kisophotofit}.
Because of the better seeing size in $K'$ than that of the $J$ band,
the ellipse fitting near the nucleus is more reliable in the $K'$ band.

We find that the position angle abruptly changes from $\sim -8^\circ$ 
(i.e., the same as the position angle of galactic disk major axis)
to $\sim -23^\circ$ at a semimajor axis of $\sim$ 8$''$ or 730 pc.
This can also be seen in the contour map in figure \ref{fig:Kcloseup},
and may suggest the existence of an oval distortion.
It is difficult to conclude the presence of a bar here, however.
Generally, the criteria for the identification of bars are
(i) the ellipticity increases as a function of the radius 
to a maximum, and then decreases to 
reveal the inclination of the disk, and (ii) the position angle is
constant over the radii where the ellipticity is rising 
(McLeod, Rieke 1995; Mulchaey et al.\ 1997),
but the highly inclined disk of NGC 5033 prevents us 
from a reliable identification of the bars.
Moreover, we find that the existence of the molecular spiral arms could also 
influence the isophotos, even in the $K'$ band.

If the observed NIR features are 2-dimensional,
we can deproject the isophotal contours by expanding figure \ref{fig:Kcloseup}
along the minor axis by a factor of (cos $i$)$^{-1}$; 
we then see the elongated distribution of the $K'$ emission
at a semi-major length of $\sim$ 15$''$. 
However, it is likely that the true distribution of low-mass stars, 
which dominate the $K'$ emission observed here, is 3-dimensional, 
and hence the ``deprojection'' of the contour maps 
gives an upper limit of the oval structures in the $K'$-band emission.

\subsubsection{$J-K'$ color}

The $K'$-band image was convolved to the same resolution as the $J$-band data,
and a $J-K'$ color map was produced.
Figure \ref{fig:colorvsCO} shows the $J-K'$ color map
and a comparison with the integrated CO intensity image.
The CO image was aligned to NIR maps postulating the NIR continuum peak
as the position of the Seyfert nucleus. 

A circumnuclear ring with a radius of $\sim$ 1.3 kpc
and two spiral arms are clearly seen
in the $J-K'$ color. 
These spirals are also clearly confirmed by the high angular resolution
NIR colors (\cite{pel99}).
We see that the ring and the two spiral arms in the $J-K'$ color map agree well 
qualitatively with the CO distribution. 
We can trace the successive CO peaks along the ring 
and two spirals seen in the $J-K'$ map.
The two spirals seen in the $J-K'$ map terminate at a radius of $\sim 30''$ 
as well as the CO map. 
Thus, the abrupt cut-off of the CO emission seen in figure 2b means that 
most of the ISM, including H\,{\footnotesize I} gas, 
is really concentrated within this radius.
This is supported by a CO mapping over the entire disk (\cite{nn01}; \cite{cur01};
\cite{won02})
and an 850 $\mu$m dust continuum map (Doi 2002, private communication).

Good correlations between CO maps and NIR color maps have been reported 
in other spiral galaxies (e.g., Sakamoto et al.\ 1995; 
Hurt et al.\ 1996; \cite{reg00}).
The correspondence between CO and $J-K'$ color is particularly significant 
in the western side of the galaxy. 
This could mean that the western part of the galaxy is the near side;
because the ISM would be distributed on a thin disk-like plane, 
in contrast to stars,
more stars on the near side are reddened than those on the far side.
This configuration indicates that the spiral arms seen 
in optical and CO images are trailing.

On the other hand,
we see a significant discrepancy between the NIR color index and
the CO distribution within the central $r \leq 5''$ region;
the color map shows a single peak at the Seyfert nucleus, 
whereas there exist two CO peaks apart from the nucleus.
It could be likely that the NIR color near the active nucleus
no longer reflects the dust extinction (and therefore the distribution of the ISM),
as in the case of the nucleus in NGC 1068 (Hurt et al.\ 1996).
In fact, HST observations of NGC 5033 at 1.6 $\mu$m band show an unresolved source
in the center of NGC 5033. It must be from the non-stellar activity,
because its flux correlates with the hard X-ray luminosity (\cite{qui01}).
The $J-K'$ index toward the nucleus is 1.21 mag in NGC 5033,
which is a redder color. Similar $J-K'$ color indices toward Seyfert nuclei
can be found in the literature (e.g., Forbes et al.\ 1992; Hunt, Giovanardi 1992;
Alonso-Herrero et al.\ 1996, 1998; \cite{pel99}).

\section{Discussion}

Our high-resolution CO and NIR observations reveal a non-axisymmetric molecular gas
distribution and perturbed gas kinematics 
in the central $\sim$ kpc region of NGC 5033.
The observed morphological features are:
(i) two strong CO peaks straddling the nucleus, separated by $\sim$ $3''$ 
or 270 pc from the nucleus,
(ii) a molecular ring with a radius of $\sim$ $14''$ or 1.3 kpc,
(iii) offset molecular ridges connecting the two CO peaks 
and the molecular ring, and 
(iv) two-armed molecular spirals that terminate abruptly
at a radius of $\sim$ $30''$ or 2.7 kpc.   
We also find kinematical features:
(v) double peaked and wide CO line profiles near the two CO peaks, and
(vi) a twist of the isovelocity contours near the systemic velocity, 
which tend to align with the position angle of the CO ridges.

In this section, we show that all of the observed molecular gas distribution 
and kinematics can be understood as a response to a {\it small} nuclear bar.
We also discuss the gravitational instabilities of molecular gas 
in the center of NGC 5033
in order to show that no starburst coexists 
in the low-luminosity type-1.5 Seyfert nucleus of NGC 5033.

\subsection{Signature of a Nuclear Bar in NGC 5033}

The observed features listed above are reminiscent of those 
in the central regions of barred spiral galaxies;
first, the two conspicuous CO peaks which appear to straddle 
the nucleus of the NGC 5033
quite resemble the ``CO twin peaks'' widely observed in the central regions 
of barred galaxies (e.g., Kenney et al.\ 1992; Kenney 1996).
Second, the elongations of CO emission from two CO peaks
mimic the dust-lane morphology of barred galaxies 
(e.g., Athanassoula 1992). 
Offset ridge structures of CO emission along the leading sides of a bar
are often observed in the central regions of barred galaxies, 
such as IC 342 (\cite{ish90}), 
NGC 1530 (Reynaud, Downes 1998; \cite{reg99}), 
and NGC 3504 (Kuno et al.\ 2000).
These morphological similarities of the CO distribution tempt us 
to consider the existence of a bar 
at a position angle of $-50^\circ \pm -10^\circ$,
because the twin peaks seen in barred galaxies
are usually oriented perpendicular to the stellar bars, 
and because CO ridges should be located at the leading sides of the bars.
The observed velocity field, where isovelocity contours deviate 
from the circular motion and tend to align with the position angle of the CO ridges, 
is also consistent with the presence of a bar.
This is because models of the gas dynamics in a barred potential 
predict that the largest velocity gradient should be seen across the bar 
(e.g., Robert et al.\ 1979); in fact, such features have been observed 
in the central regions of galaxies with bars.

In addition to these, the apparent disagreement between the P--V diagram
and the circular rotation velocities from a moment map 
seen in figure 7 could also be a suggestive of a bar. 
As mentioned in subsubsection 3.1.6, a steeper P--V than
the rotation curve can occur due to a bar
(Bureau, Athanassoula 1999; Athanassoula, Bureau 1999).

\subsection{Nuclear Bar and Dynamical Resonances}

In order to check whether an orbit resonance theory can explain the observed
gas distribution, we examined the angular velocities of the molecular gas.
The inner Lindblad resonances (ILRs) will exist wherever 
$\Omega(r)-\kappa(r)/2 = \Omega_{\rm bar}$ at some location within the corotation,
where $\Omega(r)$ is the rotation frequency at a radius of $r$
[i.e., $= v(r)/r$], 
$\Omega_{\rm bar}$ is the rotation frequency of the bar 
(the pattern speed of the bar, and is constant with radius),
and $\kappa(r)$ is the epicyclic frequency.
For a higher order resonance, the ultra harmonic resonance (UHR; 4:1 resonance)
can also occur, where $\Omega(r)-\kappa(r)/4 = \Omega_{\rm bar}$.

We think that there are two possible loci of the corotation resonance 
[CR; it occurs where $\Omega(r) = \Omega_{\rm bar}$] 
from the observed CO distribution:
one is to assume that (i) a cut-off of the molecular gas distribution corresponds to
a possible corotation radius, 
giving a possible CR radius $R_{\rm CR}  \sim 30''$ or 2.7 kpc,
and the other is (ii) to assume the CR lies near $R_{\rm CR} \sim 23''$ or 2.1 kpc,
where a local dip of CO
is seen in the radial distribution of the molecular gas
(figure \ref{fig:COradialdistribution}).
These are because barred potentials sweep gas clouds inwards from the corotation
up to ILRs; also because the gas outside the corotation tends to accumulate near the
outer Lindblad resonance (OLR),
the ISM would be depopulated in the corotation resonance.
This scenario has been successfully applied to
several galaxies, such as NGC 891 (Garc\'{\i}a-Burillo, Gu\'elin 1995) 
and NGC 253 (Houghton et al.\ 1997). 
These assumptions immediately yield possible bar pattern-speeds of 
$\Omega_{\rm bar} \sim$ 85 km s$^{\rm -1}$ kpc$^{\rm -1}$ for the first case (i) and
$\Omega_{\rm bar} \sim$ 105 km s$^{\rm -1}$ kpc$^{\rm -1}$ for the latter case (ii), 
respectively,
adopting the rotation curve discussed in subsubsection 3.1.6 
(also see figure \ref{fig:pv}).

Figure \ref{fig:resonance} shows the angular velocities as a function 
of the galactocentric radius, i.e., 
$\Omega(r) = v(r)/r$, $\Omega(r) \pm \kappa(r)/2$, and $\Omega(r) - \kappa(r)/4$,
where
$\kappa(r) = \{ 2 [ v(r)/r ] \cdot [ v(r)/r + dv(r)/dr ] \}^{0.5}$,     
as well as the proposed range of the bar pattern speeds, $\Omega_{\rm bar}$.
This resonance analysis means that the two observed CO peaks and 
the elongated CO ridges lie between the inner ILR (IILR) and outer ILR (OILR).
This situation is widely observed in the central regions of barred galaxies
(e.g., Kenney et al.\ 1992; Telesco et al.\ 1993; Kenney 1996; 
Downes et al.\ 1996; Kohno et al.\ 1999a),
although it must be treated with caution, because
non-circular motions near the center of barred galaxies can severely affect
the estimation of ILR radii (Sakamoto et al.\ 1999a).  
These CO peaks are considered to be orbit crowding zones 
where different families of gas streams, 
$x_1$ and $x_2$, would be crossing. 
This would result in shocks along the orbit crowding region, 
as suggested by theory (e.g., Athanassoula 1992).
In fact, very large velocity dispersion and/or complex CO spectra 
with two or more
peaks have been observed near the ``twin peaks'' of barred galaxies
(Kenney 1996; Kohno et al.\ 1999a; \cite{sor00}; \cite{kod02}), 
adding another bit of evidence for a bar-driven gas motion.
Figure \ref{fig:resonance} also shows that the 1.3 kpc ring feature may correspond to
the UHR if $\Omega_{\rm bar} \sim$ 105 km s$^{\rm -1}$ kpc$^{\rm -1}$ 
(i.e., the assumption (ii)).
Nuclear rings possibly explained by UHR are observed in 
barred spiral galaxies, such as NGC 253 (\cite{sor00}) and NGC 5005 (\cite{sak00});
we therefore suggest that the larger $\Omega_{\rm bar}$ 
(i.e., $\sim$ 105 km s$^{\rm -1}$ kpc$^{\rm -1}$)
may be more plausible.
Considering the uncertainties of the rotation velocities,
we conclude here that $R_{\rm CR}$ is in the range of about 2.1 -- 2.7 kpc.

Given the $R_{\rm CR} =$ 2.1 -- 2.7 kpc, 
this would also impose an upper limit on the bar length. 
In early type galaxies, it is now widely accepted that
the corotation radius lies near to the bar end; that is,
$R_{\rm CR}$ is $\sim 1.2 \pm 0.2$ $R_{\rm bar}$ 
(Athanassoula 1992; Elmegreen 1996).
On the other hand, a numerical simulation by Combes and Elmegreen (1993)
suggested that bars in late-type galaxies may have their corotation radii
well outside the bar ends; possibly, $R_{\rm CR}$ is $\sim$ 2 $R_{\rm bar}$
(see also \cite{kun00}).
Consequently, the possible bar radius (semi-major radius) 
in NGC 5033 would be in the range 
of about 0.5 -- 1.2 $R_{\rm CR}$, i.e., 1.1 -- 3.2 kpc,
and probably near 0.5 $R_{\rm CR}$ or 1.1 -- 1.4 kpc,
as in the case of the bars in other late-type spirals,
such as M 100 (Garc\'{\i}a-Burillo et al.\ 1994; Sempre et al.\ 1995; 
Elmegreen et al.\ 1992).

The possible bar radius of $\sim$ 1.1 -- 1.4 kpc 
is also implied from the observed radial distribution of CO emission; 
recent CO observations of barred galaxies
demonstrate the existence of a ``secondary peak'' of the ISM near the bar-ends
(Kenney, Load 1991; Nakai 1992; Downes et al.\ 1996; \cite{nis01}), and
the second enhancement of CO emission at a radius of $\sim$ 1.3 kpc seen 
in the CO radial distribution
of figure 4 may suggest that a plausible bar radius in NGC 5033 would
indeed be $\sim$ 1.1 -- 1.4 kpc.

We should note that NGC 5033 has several rings 
in the outer disk region ($r >$ a few arcmin),
as can be seen in the H\,{\footnotesize I} maps (\cite{the97}).
These outer rings seem to be difficult to explain
by only the small nuclear bar proposed here,
because the OLR of the nuclear bar would be located at about $r \sim 3.7$ kpc
or $\sim 40''$, which is much more inward than the outer H\,{\footnotesize I} rings.
However, it is possible to understand the whole structure
if NGC 5033 has multiple bars; 
S. Ishizuki (2002, private communication) proposed that
there are three bars in the disk of NGC 5033
based on the deprojected H\,{\footnotesize I} map (\cite{the97}), 
assuming that the ILR radii of the outer bars correspond to
the corotation radii of the inner bars
(\cite{fm93}).

In summary, both the distribution and kinematics of the molecular gas
in the central region of NGC 5033 can be well understood as the response
to a bar. We propose that a small nuclear bar lies along 
a position angle of $-50^\circ \pm -10^\circ$ 
and a possible semi-major radius of $R_{\rm bar} \sim$ 1.1 -- 1.4 kpc
($12''$ -- $15''$). The pattern speed of the nuclear bar is
estimated to be 85 -- 105 km s$^{-1}$ pc$^{-1}$.
Figure \ref{fig:faceonview} displays our proposed interpretation 
of the gas distribution
in the center of the ``non-barred'' Seyfert galaxy NGC 5033.
This small nuclear bar would sweep the molecular gas 
inside the corotation radius and channel it 
into the central region of the galaxy.

Our results also demonstrate that high-resolution CO imaging-spectroscopy
is useful to search for a bar even in highly inclined systems where 
isophoto ellipse fitting techniques are not applicable.

\subsection{Role of a Nuclear Bar}

The fact that $\sim$ 20$''$ resolution H\,{\footnotesize I} observations 
show little evidence for a bar in NGC 5033 (Thean et al.\ 1997) would suggest that
a large-scale nonaxisymmetry of underlying potential is not required
for Seyfert activity, and that high-resolution observations of the ISM 
are crucial for studying the fueling mechanism of Seyfert galaxies.
As pointed out by McLeod and Rieke (1995),
the critical elements of the fuel supply need not be visible
on the {\it large scale} in the host galaxy.
In fact, the possible bar length in NGC 5033 is quite small; 
the semi-major radius of $\sim$ 1.1 -- 1.4 kpc
corresponds to only $\sim 0.02 \times D_{25}$.
We should also note that 
a very weak bar can play a significant role on the ISM.
Numerical simulations have predicted that the molecular gas distribution 
and kinematics
severely affected even in a potential with a very weak asymmetry, 
a few \% or less in density (\cite{wh92}; Jungwiert, Palous 1996),
which might not be easy to detect, even in the NIR bands.
Hence, it seems that the apparent lack of a clear and strong bar/oval structure 
in the stellar surface density traced with NIR continuum emission 
does not immediately 
mean that the ISM in such galaxies is not driven into the central regions
by a bar.

Nevertheless, it appears to be clear that the presence of a small-scale
bar is {\it not} a {\it sufficient} condition for Seyfert activity,
though it could be a necessary condition. 
As described in the previous sections, bar-driven gas structures, 
such as the ``twin-peaks'' seen in NGC 5033, are widely observed 
in many barred spiral galaxies, 
but not every barred galaxies hosts an active nucleus. 
Additional mechanisms, such as self-gravity of ISM
(e.g., \cite{wh92,hs94}) or a ``nuclear ILR'' induced by
the central super massive black holes (Fukuda et al.\ 1998, 2000),
would be necessary to drive ISM into the central engine.

It should also be noteworthy that 
a very small amount of gas is sufficient to sustain
low-level Seyfert activities.
The required mass accretion rates, $\dot{M}$, in order to feed
a given luminosity, $L_{\rm bol}$, are expected to be
\begin{equation}
\dot{M}
= \frac{L_{\rm bol}}{\eta c^2}
= 2.0 \times 10^{-2} \left( \frac{\eta}{0.1} \right)^{-1} 
             \left( \frac{L_{\rm bol}}{10^{44}\ \mbox{erg s$^{-1}$}} \right)
             M_\odot \mbox{yr$^{-1}$} \mbox{,}
\end{equation}
(\cite{ree84}) for a typical luminosity of Seyfert galaxies,
where $\eta$ is the accretion efficiency.
Because NGC 5033 is a low-luminosity Seyfert
($L_{\rm 2 - 10 keV} = 2.3 \times 10^{41}$ erg s$^{-1}$, \cite{tera99}; 
see also Koratkar et al.\ 1995), the required gas mass is only $\sim 10^5 M_\odot$
to feed the monster over a time of $\sim 10^9$ yr.
This may explain why not all Seyfert hosts are barred; 
a high $\dot{M}$, which can be achieved by a bar, 
is not always required for low-luminosity AGNs
(Ho et al.\ 1997c).

\subsection{Star Formation and Self Gravity of Molecular Gas in Seyfert Nuclei}

A massive nuclear starburst is 
often associated with Seyfert galaxies.
In particular, a large fraction (about 50\%, \cite{sb01}) 
of type-2 Seyfert nuclei are reported to coexist with a nuclear starburst
(e.g., \cite{hec97}; \cite{mai98}; \cite{gd98}; \cite{cf01}; \cite{ima02}).
These nuclear starbursts can be in the dense obscuring torus (e.g., \cite{cft95}),
which would play an important role to form a geometrically thick torus
via energy feedback from supernovae (\cite{fab98,wn02}). 

In order to address the nuclear star formation in the type-1.5 Seyfert galaxy NGC 5033,
we examined the gas mass fraction to the dynamical mass, 
$M_{\rm gas}$/$M_{\rm dyn}$, in the center of NGC 5033. 
In galaxies with nuclear starburst/star formation,
$M_{\rm gas}$/$M_{\rm dyn}$ often exceeds 10\% 
(e.g., \cite{ken93}; Sakamoto et al.\ 1999b).
Such a high gas fraction results in the unstable states of molecular gas,
and eventually massive star formation occurs there (e.g., \cite{elm94}).
Therefore, the $M_{\rm gas}$/$M_{\rm dyn}$ values 
can be a rough diagnostic of starbursts
in the centers of galaxies.

The dynamical mass within a given radius $r$ is calculated as
\begin{equation}
M_{\rm dyn} = \frac{r v^2(r)}{G}
= 2.3 \times 10^5  \left( \frac{r}{\mbox{kpc}} \right) 
                   \left[ \frac{v(r)}{\mbox{km s$^{\rm -1}$}} \right]^2 M_\odot \mbox{,}
\end{equation}
where $v(r)$ is a circular rotation velocity at a galactocentric radius of $r$, 
assuming a spherical mass distribution.
The dynamical mass within a radius of 30$''$ or 2.7 kpc, 
the outer boundary of the CO emission, 
was then calculated to be 3.0 $\times$ 10$^{\rm 10}$ $M_\odot$,
and the ratio of dynamical mass to the total molecular gas mass 
$M_{\rm gas}/M_{\rm dyn}$ is 0.081 within $r < 2.7$ kpc. 
The gas mass ratio, $M_{\rm gas}/M_{\rm dyn}$,
remains almost constant over the molecular disk;
the $M_{\rm gas}/M_{\rm dyn}$ within $r<1.4$ kpc ($15''$), 
which corresponds to the radius of the molecular ring, is found to be 0.063.
It drops to 0.006 in the very center ($r<2''$ or 180 pc), however.

This $M_{\rm gas}/M_{\rm dyn}$ of 0.006 is very small compared with that
in nuclear starburst galaxies, where the ratio often exceeds 0.1
(e.g., Sakamoto et al.\ 1999b).
We therefore suggest that the molecular gas 
in the central $r<200$ pc region of NGC 5033
may not be self-gravitating, in contrast to the situation
in nuclear starburst galaxies.
This implies that no nuclear starburst coexists 
in the center of the type-1.5 Seyfert galaxy NGC 5033,
in accord with high-resolution NIR spectroscopic studies
(\cite{iva00}; \cite{ima02}), though extended star-forming regions exist
in the disk of NGC 5033 (e.g., Clavel et al.\ 2000; Wang, Blitz 2002).
Further analysis of high-resolution molecular gas data in type-1 Seyferts
and a comparison with that in type-2s
will help in the study of the role of nuclear starbursts in the AGN phenomenon.

The uncertainty of the derived $M_{\rm gas}/M_{\rm dyn}$ values could be
large near the center, because the circular rotation velocity adopted here 
can be affected by the presence of non-circular motions in the central
region of NGC 5033.
Nevertheless, the $M_{\rm gas}/M_{\rm dyn}$ values in the 
center of NGC 5033 can be very small
because the molecular gas mass near the center of NGC 5033 may be 
overestimated;
the conversion factors in the central regions of galaxies
(including the Galactic Center region)
tend to be {\it smaller} than those in the Galactic disk GMCs
by a factor of 3 -- 10 or more (e.g., Nakai, Kuno 1995; Solomon et al.\ 1997; 
Oka et al.\ 1998; \cite{dah98}; \cite{reg00}),
which would result in a further decreasing of $M_{\rm gas}/M_{\rm dyn}$ 
in the nucleus of NGC 5033.

\section{Conclusions}
 
Our high-resolution aperture synthesis CO(1--0) observations
and NIR broad-band photometry were performed
toward the central region of the ``non-barred'' Seyfert galaxy NGC 5033.  
The results of the observations are summarized as follows:

\begin{enumerate}

\item Our high-resolution CO images reveal a perturbed distribution 
of molecular gas in the center of the ``non-barred'' Seyfert galaxy NGC 5033; 
we find 
(i) two strong CO peaks straddling the nucleus, separated by $\sim$ $3''$ 
or 270 pc from the nucleus,
(ii) a molecular ring with a radius of $\sim$ $14''$ or 1.3 kpc, and
(iii) offset molecular ridges connecting the two CO peaks 
and the molecular ring.  
These are the characteristic gaseous features that are widely
observed in the centers of barred galaxies.

\item We find that broad ($\geq$ 200 km s$^{\rm -1}$ in FWHM) and
double-peaked profiles are seen near the two CO peaks,
which would support the idea that the two observed CO concentrations
are the orbit crowding regions where different families of gas orbits
($x_1$ and $x_2$) in the bar meet there.
The observed velocity field also suggests the presence of a bar in NGC 5033.

\item Outside the ring, we find two armed spiral structures, 
which abruptly terminate at the radius of $\sim$ $30''$ or 2.7 kpc, 
showing an excellent agreement to the $J-K'$ color index map.
However, in the very center of the active nucleus, 
the $J-K'$ color does not reflect the distribution of ISM.

\item Based on the resonance analysis of the molecular gas,
we propose that the gas distribution and kinematics in the central
a few kpc region of NGC 5033 are governed by a small nuclear bar.
The possible bar properties, i.e., the position angle, semi-major radius,
and pattern speed, are estimated to be about $-50^\circ \pm 10^\circ$,
$\sim$ 1.1 -- 1.4 kpc ($12''$ -- $15''$), 
and $\sim$ 85 -- 105 km s$^{-1}$ pc$^{-1}$, respectively.
The parameter of the nuclear bar proposed here can account for 
all of the observed CO features with consistency.

\item Although we could not confirm the presence
of a nuclear bar from NIR photometry 
due to the high inclination angle of NGC 5033,
we did obtain an upper limit of the bar radius, $\sim 20''$. 
This is consistent with the size of the bar
deduced from the CO data.
These results demonstrate that high-resolution CO imaging-spectroscopy
is useful to search for a bar even in highly inclined systems where 
isophoto fitting techniques are not applicable.

\item The fraction of molecular gas mass to the dynamical mass,
$M_{\rm gas}/M_{\rm dyn}$, is very small 
($\leq1$ \% within $r<2''$ or 180 pc).
This would suggest that the molecular gas accumulated in this region
is not self-gravitating, 
in contrast to nuclear starburst/star-forming galaxies,
implying no nuclear starburst coexists 
in the type-1.5 Seyfert galaxy NGC 5033.
It should be noted that the $M_{\rm gas}/M_{\rm dyn}$ values 
near the center could contain large uncertainties due to a strong
non-circular motion in the central region of NGC 5033, however.

\end{enumerate}

\bigskip

We would like to thank the referee for helpful comments
that improved this paper.
We are deeply indebted to the NRO staff for operating the telescopes
and continuous efforts to improve the performance of the instruments.
We are grateful to A. Mori for his extensive help during the observations
at Okayama Astronomy Observatory, S. Okumura and T. Minezaki
for their advice and comments 
on the data reduction and interpretation of our NIR data,
and T. Kamazaki for providing us with 
his useful software packages to make figures in this paper.
Nobeyama Radio Observatory (NRO) and Okayama Astrophysical Observatory (OAO) 
are a branch of the National Astronomical Observatory, an interuniversity
research institute, operated by the Ministry of Education, Culture, Sports, 
Science and Technology.

\clearpage

\begin{figure}
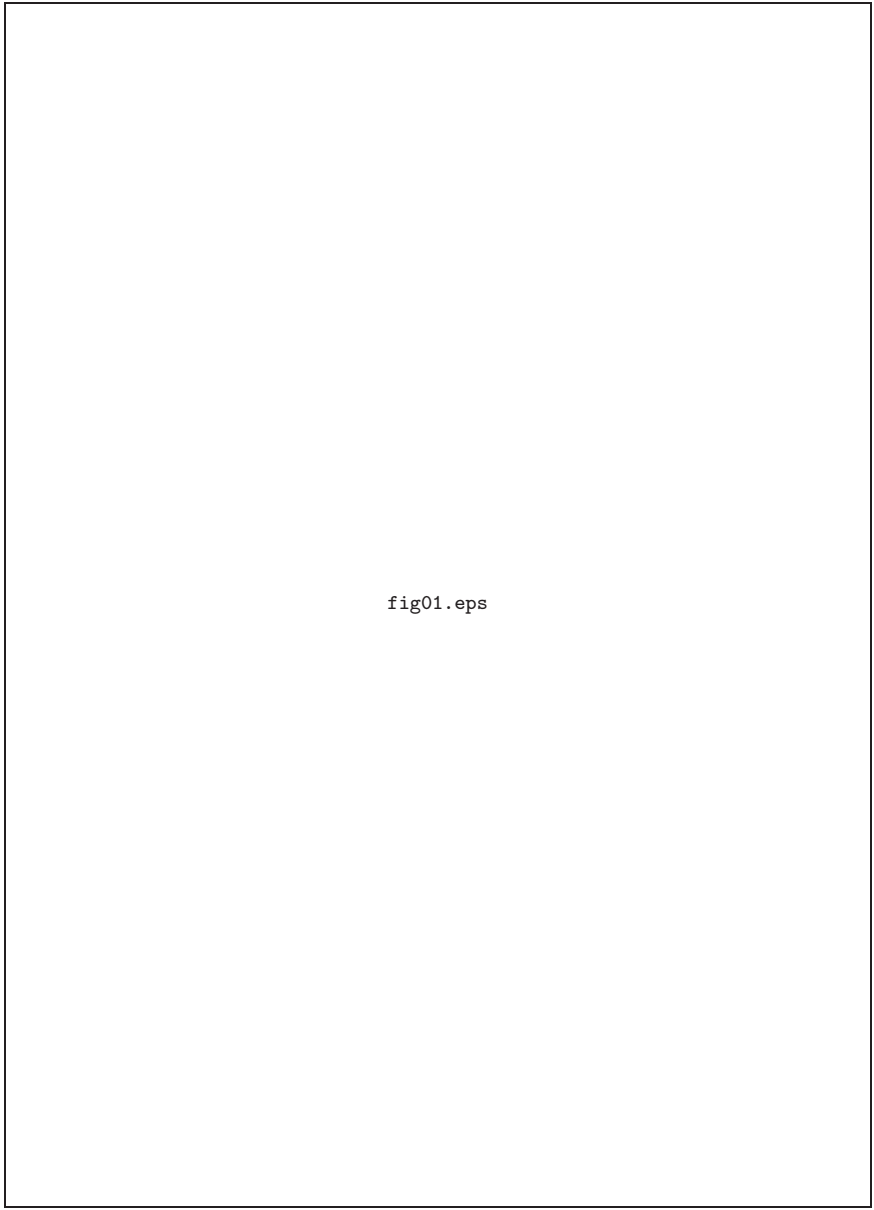

  \begin{center}
    \FigureFile(115mm,160mm){fig01.eps}
  \end{center}
\caption{
Channel maps of CO(1--0) emission from the
central $45'' \times 75''$ region 
(4.1 kpc $\times$ 6.8 kpc at $D$ = 18.7 Mpc) of NGC 5033.
Velocity channels with a width of 19.55 km s$^{\rm -1}$
are displayed at an interval of 9.78 km s$^{\rm -1}$.
The central velocity ($V_{\rm LSR}$ in km s$^{\rm -1}$)
of each channel is labeled.
The size of the synthesized beam is 
$3.\hspace{-2pt}''9 \times 3.\hspace{-2pt}''6$ (HPBW), shown as an ellipse.
The contour levels are $-4$, $-2$, 2, $\cdots$, $12 \sigma$,
where 1 $\sigma$ = 30 mJy beam$^{\rm -1}$ or 200 mK in $T_{\rm b}$.
The large cross indicates the position of the 6 cm radio continuum peak,
and the small cross shows the pointing center.
The circle represents the field of view ($65''$). 
The attenuation due to the primary beam pattern
is not corrected in these maps.
}
\label{fig:chanmap}
\end{figure}

\addtocounter{figure}{-1}
\begin{figure*}
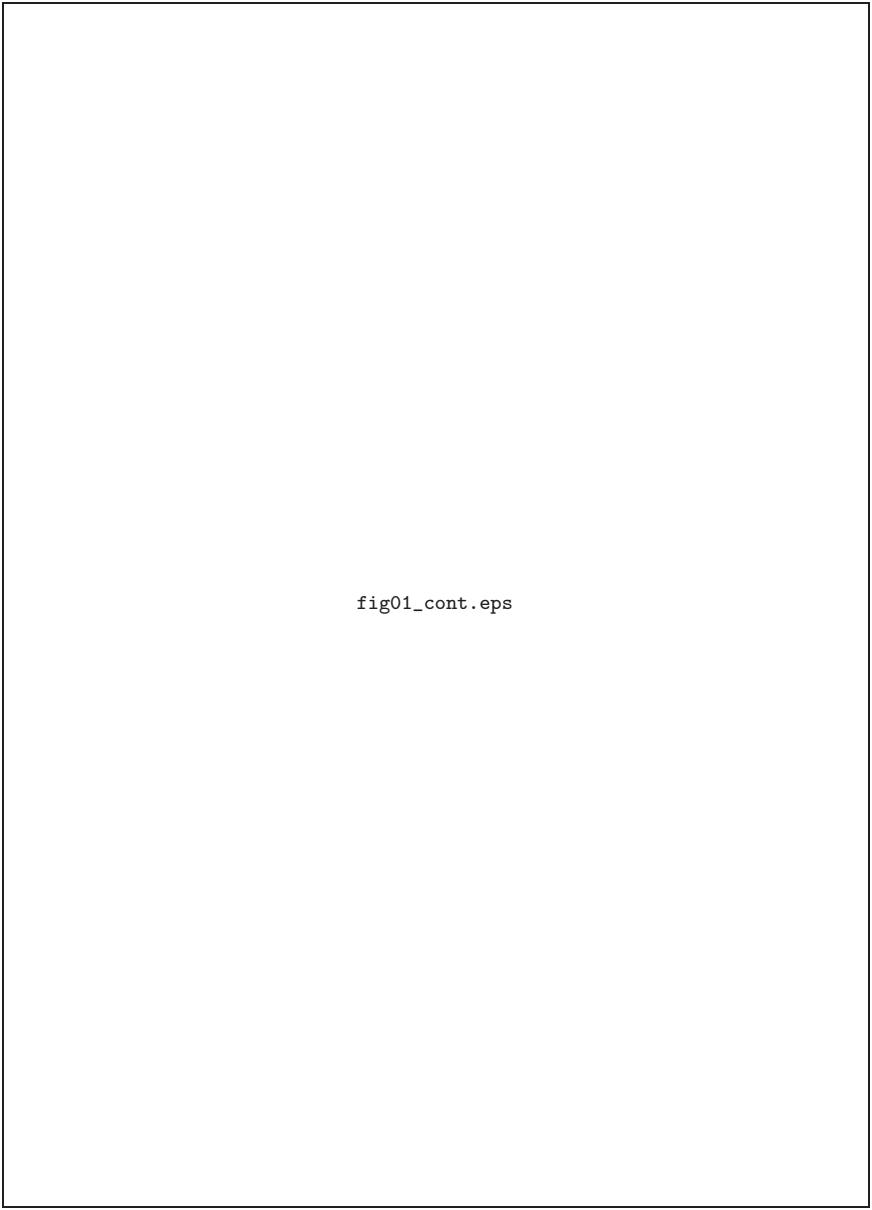

 \begin{center}
  \FigureFile(115mm,160mm){fig01_cont.eps}
 \end{center}
\caption{(Continued)}
\label{fig:chanmap2}
\end{figure*}

\begin{figure}
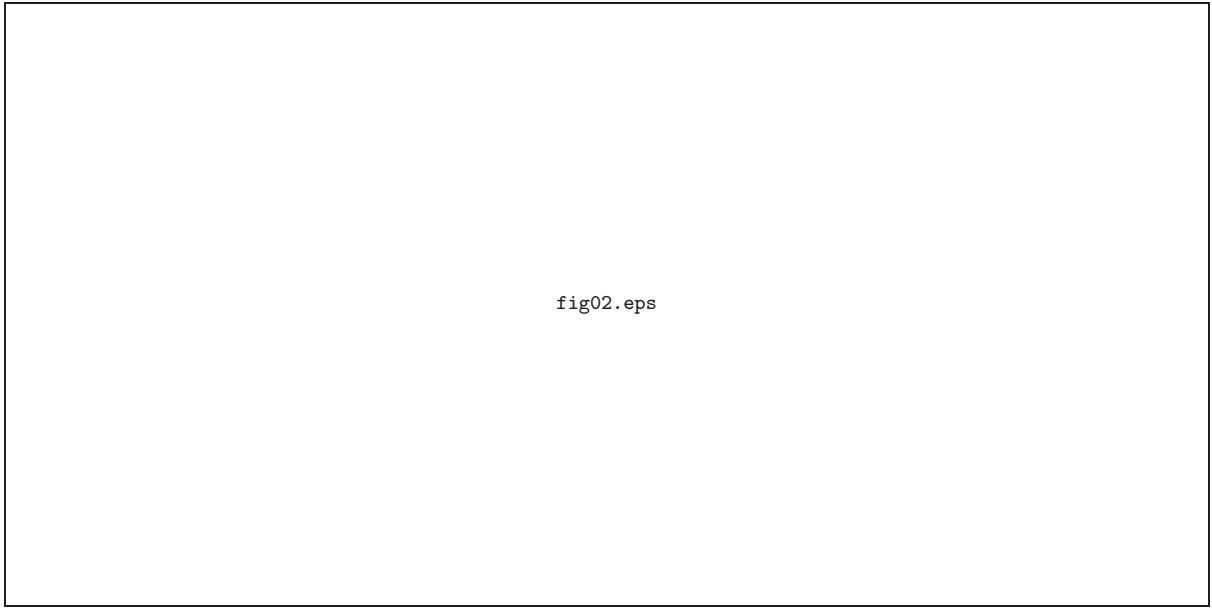

  \begin{center}
    \FigureFile(160mm,80mm){fig02.eps}
  \end{center}
\caption{
(a) Optical image of NGC 5033 from the Digitized Sky Survey, showing
a $3' \times 6'$ (16 kpc $\times$ 33 kpc at $D = 18.7$ Mpc)
field around the center of NGC 5033.  
(b) Integrated intensity map of the CO(1--0) emission over a velocity
range from $V_{\rm LSR}$ = 651.5 to 1119.4 km s$^{\rm -1}$ 
in the central $45'' \times 75''$ (4.1 kpc $\times$ 6.8 kpc)
region of NGC 5033. 
The synthesized beam, 3$.\hspace{-2pt}''$9 $\times$ 3$.\hspace{-2pt}''$6 
with a $P.A.$ of 6$^\circ$,
is shown as an ellipse in a box.
The contour levels are 1.5, 3, 4.5, $\cdots$, 13.5 $\sigma$,
where $1 \sigma$ = 2.0 Jy beam$^{\rm -1}$ km s$^{\rm -1}$ 
or 13 K km s$^{\rm -1}$ in $T_{\rm b}$. 
The $1 \sigma$ level corresponds to the face-on molecular gas (including
He and heavier elements) surface density 
$\Sigma_{\rm gas}$ = 23 $M_\odot$ pc$^{-2}$.
The large cross indicates the position of the 6 cm radio continuum peak 
[$\alpha$ (B1950) = 
13$^{\rm h}$11$^{\rm m}$09$^{\rm s}\hspace{-5pt}.\hspace{2pt}$23,
 $\delta$ (B1950) = +36$^\circ$51$'$30$.\hspace{-2pt}''$27],
and the small cross indicates the phase center of the observations.
The circle represents the field of view (HPBW of the primary beam, $65''$).
Attenuation due to the primary beam pattern
is corrected in this map.
Note that the peak of CO emission is not located at the Seyfert nucleus.
(c) Intensity-weighted mean velocity map of CO emission
in the central 45$''$ $\times$ 75$''$ region (4.1 kpc $\times$ 6.8 kpc at
$D$ = 18.7 Mpc) of NGC 5033.
The contour interval is 20 km s$^{-1}$, and the systemic velocity,
$V_{\rm LSR} = 882$ km s$^{-1}$, is indicated by a black contour.
}\label{fig:COmaps}
\end{figure}

\begin{figure}
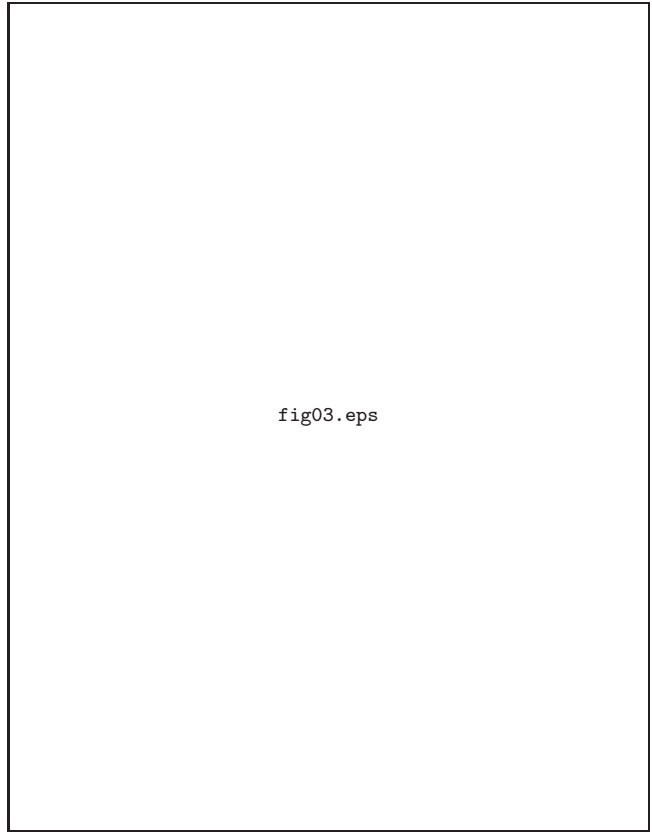

  \begin{center}
    \FigureFile(85mm,110mm){fig03.eps}
  \end{center}
\caption{
Integrated intensity map of the CO(1--0) emission over a velocity
range from $V_{\rm LSR}$ = 651.5 to 1119.4 km s$^{\rm -1}$ 
in the central 45$''$ $\times$ 75$''$ region (4.1 kpc $\times$ 6.8 kpc at
$D$ = 18.7 Mpc) of NGC 5033. Part of the molecular ring and spirals
are indicated as thick lines.
}\label{fig:COringandspiral}
\end{figure}

\begin{figure}
  \begin{center}
    \FigureFile(160mm,80mm){fig04.eps}
  \end{center}
\caption{
CO spectra from the central 21$''$ $\times$ 21$''$ region of NGC 5033.
The spectra were measured from the NMA data cube 
at 7 $\times$ 7 points on a 3$''$ spacing grid
with respect to the major axis ($P.A.$ = -8$^\circ$).
The grid positions are indicated
as crosses on the integrated intensity map. 
The contour levels of the CO map are 2, 5, 8, and 11 $\sigma$
where 1 $\sigma$ = 2.0 Jy beam$^{\rm -1}$ km s$^{\rm -1}$
or 13 K km s$^{\rm -1}$ in $T_{\rm b}$.
This CO map is rotated by 8$^\circ$ counter clockwise.
The velocity bin of the spectra is 9.78 km s$^{\rm -1}$.
The primary beam attenuation is corrected.
The noise level near the central position is 30 mJy beam$^{-1}$
or 200 mK in $T_{\rm b}$.
Very broad and double-peaked CO profiles are evident 
near the (0, 3) and (0, -3) positions, which correspond to the
two CO peaks in the map.
}\label{fig:COspectra}
\end{figure}

\begin{figure}
  \begin{center}
    \FigureFile(85mm,80mm){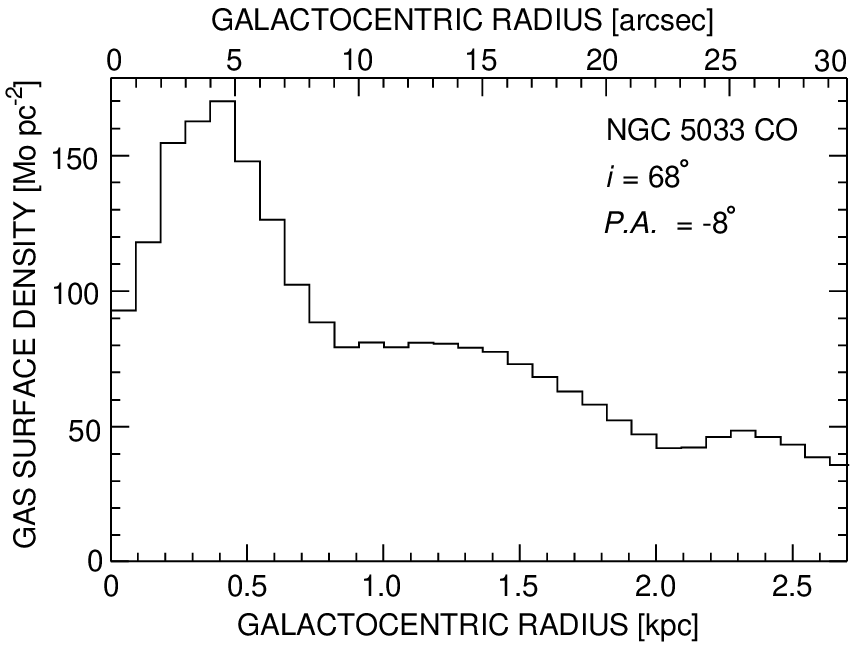}
  \end{center}
\caption{
Radial distribution of the molecular gas surface density in NGC 5033.
The CO intensities corrected to face-on were azimuthally averaged 
over the successive annuli with 1$''$ = 90.7 pc.
$X_{\rm CO} = 1.8 \times 10^{20}$ cm$^{-2}$ (K km s$^{-1}$)$^{-1}$ was adopted
to compute the gas surface density.
This surface density includes heavy elements 
as 1.36 $\times$ $\Sigma_{\rm H_2}$.
The maximum of gas surface density is located at a radius of $\sim$ 400 pc,
not at the center. The second peak is also seen at $r \sim 14''$ or 1.3 kpc.
}\label{fig:COradialdistribution}
\end{figure}

\begin{figure}
  \begin{center}
    \FigureFile(85mm,80mm){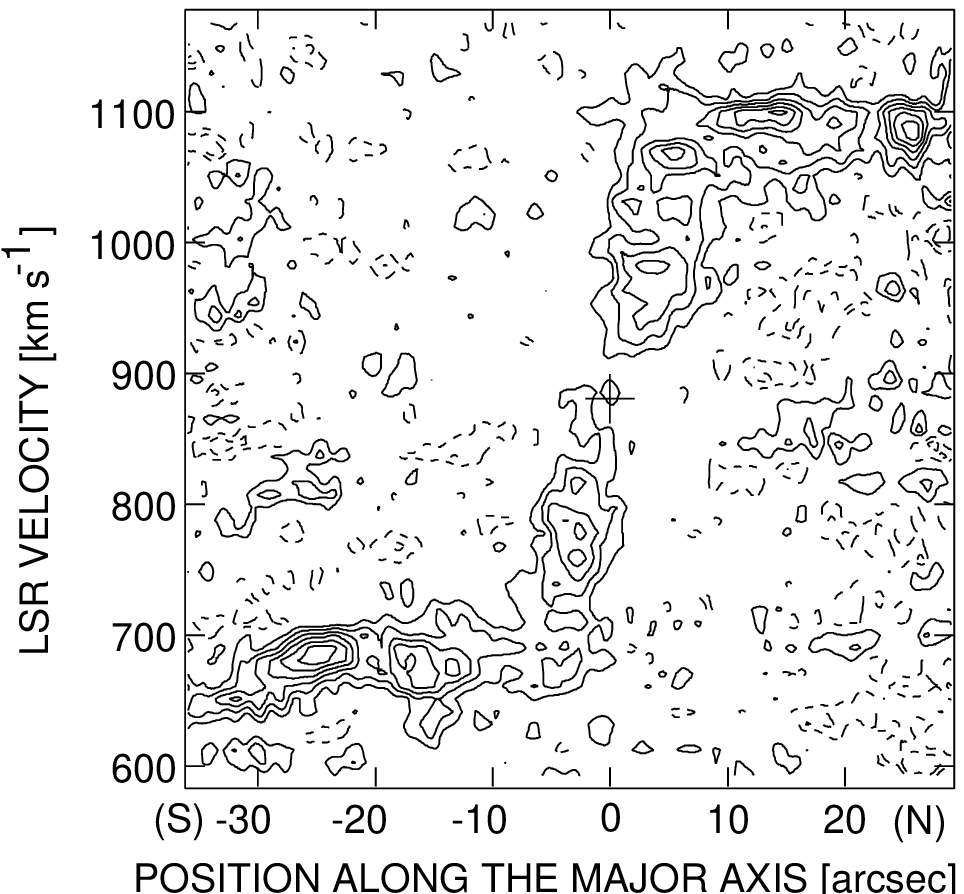}
  \end{center}
\caption{
Position -- velocity map (P--V map) of the CO emission
along the major axis of NGC 5033 ($P.A.$ = -8$^\circ$).
Contour levels are -3, -1.5, 1.5, $\cdots$, and 9 $\sigma$,
where 1 $\sigma$ = 30 mJy beam$^{\rm -1}$ or 200 mK in $T_{\rm b}$.
Negative contours are dashed.
The primary beam correction is applied.
The cross indicates the kinematical center
[$\alpha$ (B1950) = 
13$^{\rm h}$11$^{\rm m}$09$^{\rm s}\hspace{-5pt}.\hspace{2pt}$23,
 $\delta$ (B1950) = +36$^\circ$51$'$29$.\hspace{-2pt}''$9,
 and $V_{\rm sys}$ = 882 km s$^{-1}$].
}\label{fig:pv}
\end{figure}

\begin{figure}
  \begin{center}
    \FigureFile(85mm,80mm){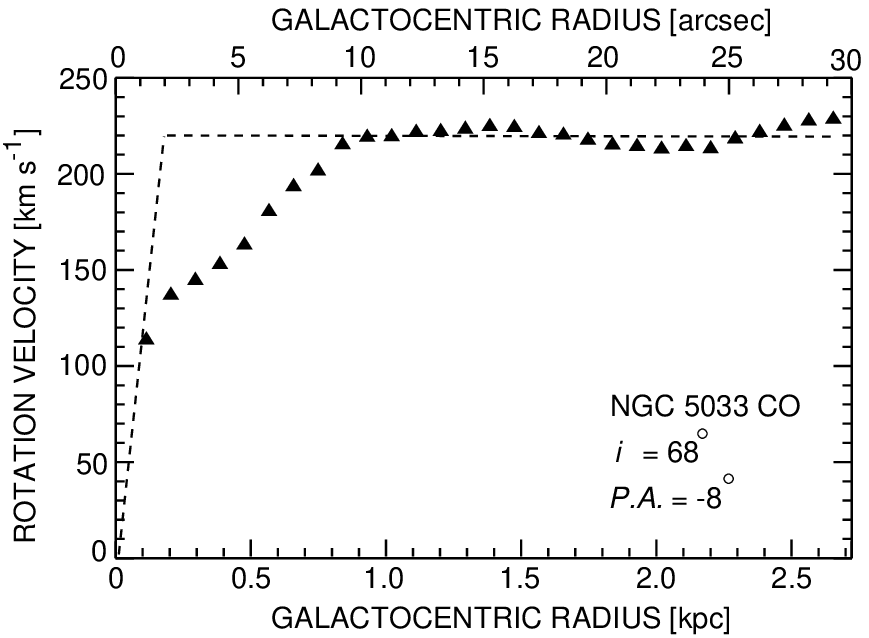}
  \end{center}
\caption{
Circular rotation curve of NGC 5033. This is derived from the
intensity-weighted mean velocity map of the CO emission 
within $\pm$ 5$^\circ$ of the major axis.
Inclination (68$^\circ$) is corrected.
Note that a gradual rise in the rotation velocity near the center ($r<10''$)
could be due to a smearing of the observing beam and multiple
velocity components near the center due to a bar,
and the true rotation velocity must be larger there,
as can be seen in the P--V diagram (figure 6).
In the following analysis, 
we assume a constant rotation velocity of
220 km s$^{-1}$ for $r>2''$,
and a rigid-body rotation of $\Omega = 1.2$ km s$^{-1}$ pc$^{-1}$
for $r<2''$, as indicated by dashed lines.
}\label{fig:rotcurve}
\end{figure}

\begin{figure}
  \begin{center}
    \FigureFile(85mm,110mm){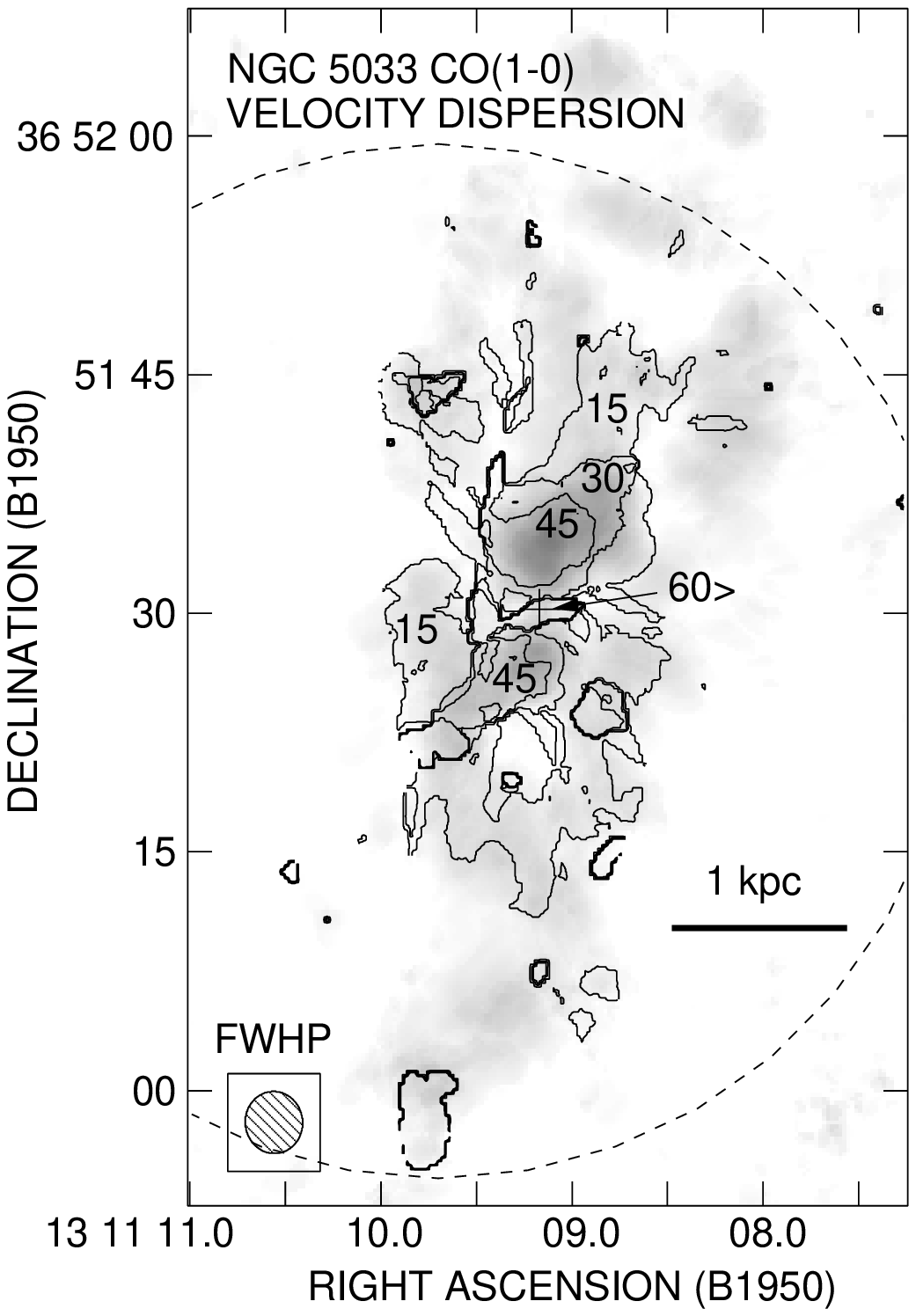}
  \end{center}
\caption{
Intensity-weighted velocity dispersion map of the CO emission.
The contour interval is 15 km s$^{\rm -1}$.
This map is computed by calculating the second-order moment of
intensity at each point from the CO cube.
The cross shows the position of the 6 cm radio nucleus, and
the dashed circle represents the field of view (HPBW of the primary beam),
that is 65$''$ diameter. 
Large velocity dispersions near the two CO peaks, 
$\sigma_v > 40$ km s$^{-1}$, are evident.
This map contains both the intrinsic velocity dispersion of gas
and the gradient of rotation velocity in the observed beam.
}\label{fig:dispersion}
\end{figure}

\begin{figure}
  \begin{center}
    \FigureFile(160mm,80mm){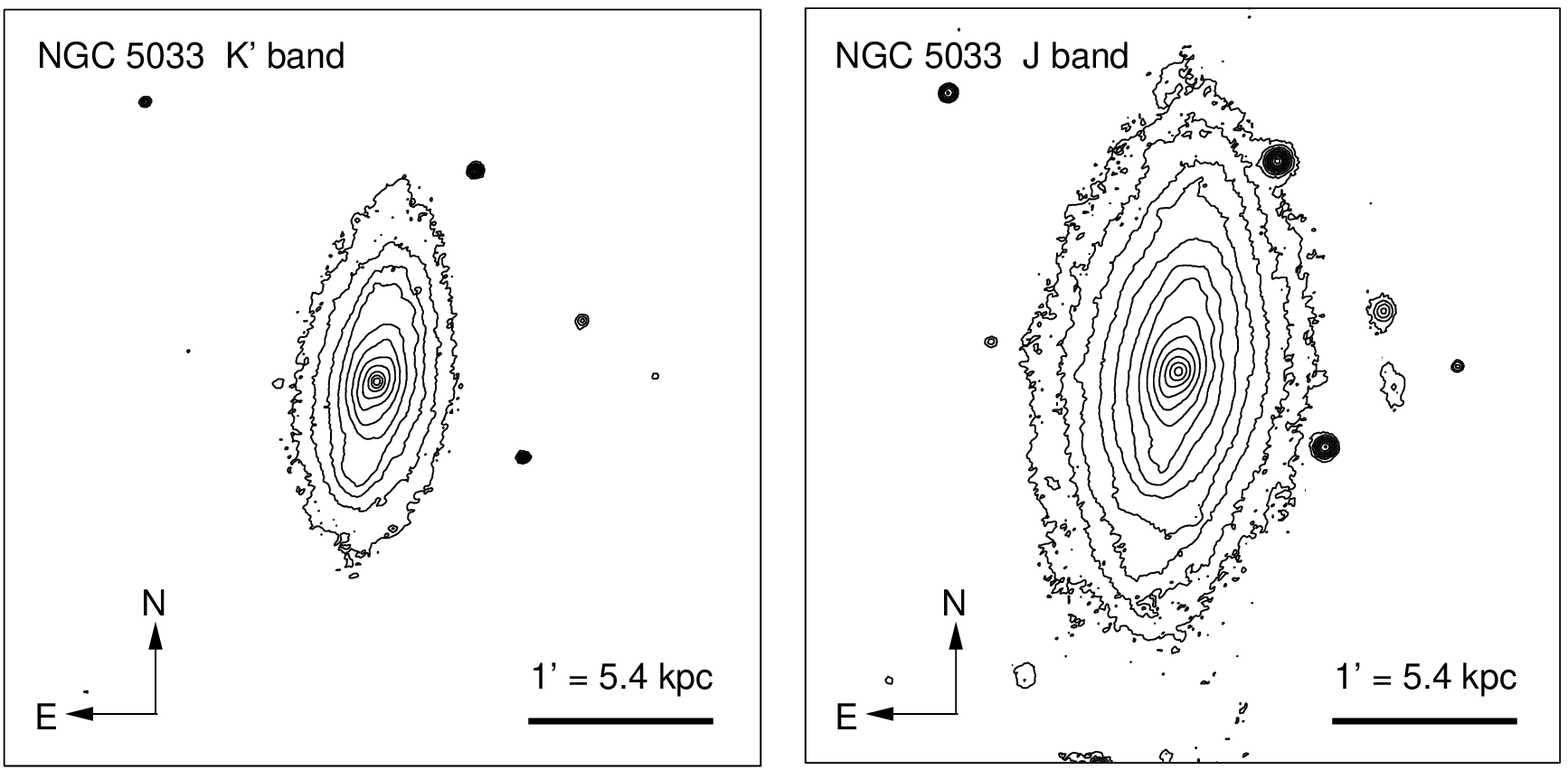}
  \end{center}
\caption{
Near-infrared images of NGC 5033 in a field of 4$.\hspace{-2pt}'$1 $\times$
4$.\hspace{-2pt}'$1 (22.3 kpc square). 
The contour levels are 18.0, 17.5, 17.0, $\cdots$, 13.0 mag arcsec$^{-2}$ 
for $K'$ the band,
and 20.5, 20.0, 19.5, $\cdots$, 14.5 mag arcsec$^{-1}$ for the $J$ band.
The horizontal bars indicate the 1$'$ or 5.44 kpc at $D$ = 18.7 Mpc.
The seeing size is about 1$.\hspace{-2pt}''$7 for the $K'$ band 
and 3$.\hspace{-2pt}''$2 for the $J$ band, respectively.
}\label{fig:JandK}
\end{figure}

\begin{figure}
  \begin{center}
    \FigureFile(85mm,80mm){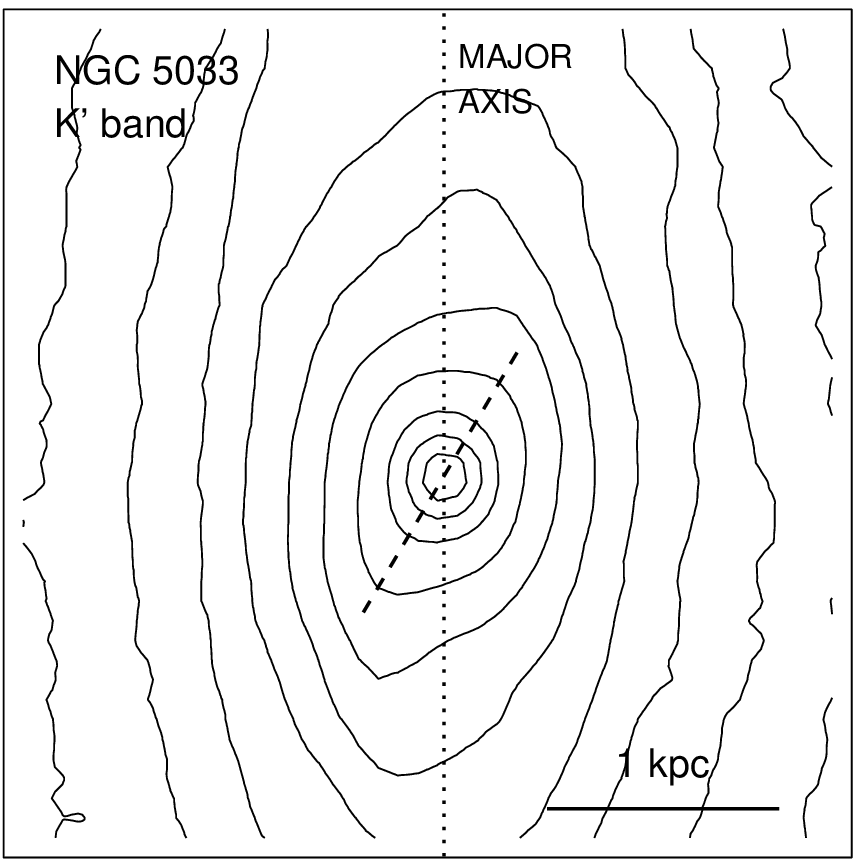}
  \end{center}
\caption{
Central 40$''$ $\times$ 40$''$ area in $K'$ band.
The contour levels are same as in figure 9.
This map is rotated by 8$^\circ$ counter clockwise.
The major axis is denoted as the dotted line.
There exists a possible shift of the isophotal contours
near the center, as indicated by the short dashed line.
}\label{fig:Kcloseup}
\end{figure}

\begin{figure}
  \begin{center}
    \FigureFile(100mm,130mm){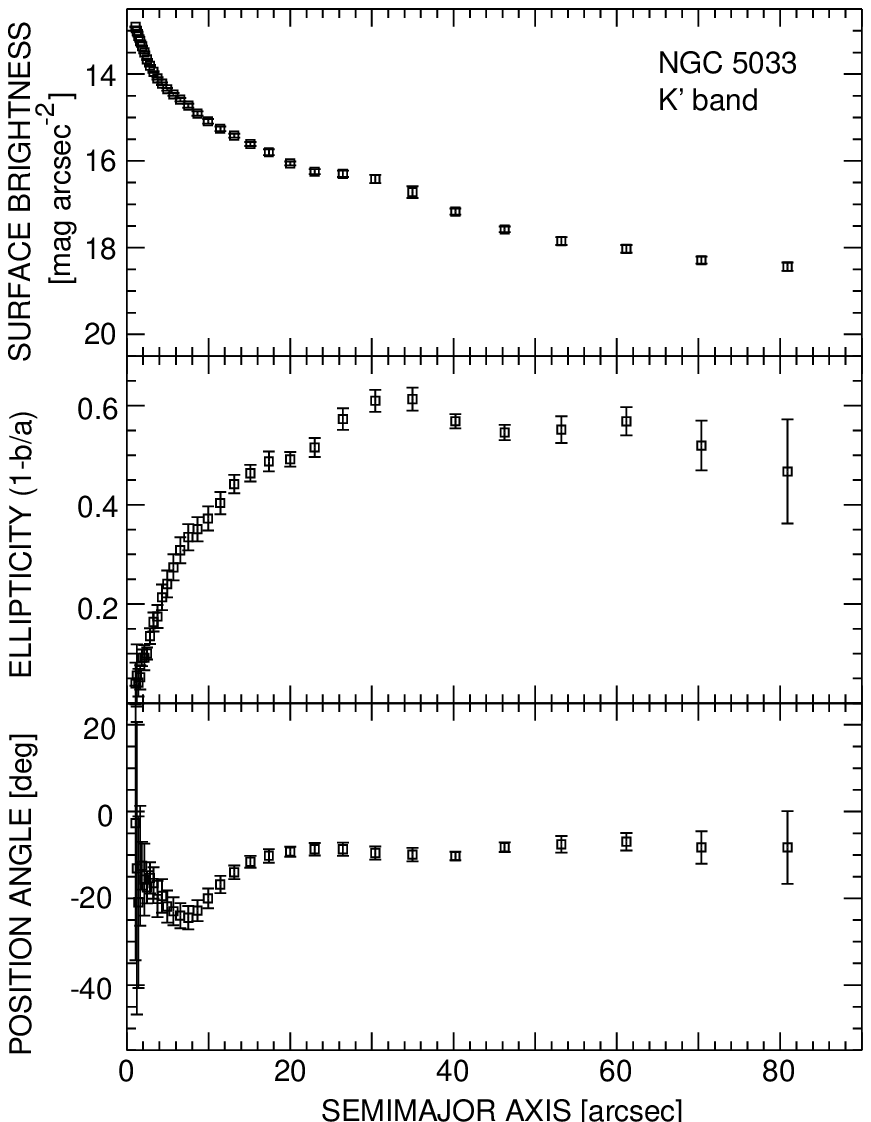}
  \end{center}
\caption{
Mean surface brightness, 
ellipticity ($1-b/a$), and position angle (North is $0^\circ$, 
and West is $-90^\circ$) as a function of the semimajor axis radius 
of elliptical isophotos fitting 
for the $K'$ band image of NGC 5033. 
The error bars are $\pm$ 1 $\sigma$.
}\label{fig:Kisophotofit}
\end{figure}

\begin{figure}
  \begin{center}
    \FigureFile(160mm,80mm){fig12.eps}
  \end{center}
\caption{
$J-K'$ color index image (left), and the comparison of the color index 
with the integrated CO distribution (right) 
in the central 2$.\hspace{-2pt}'$05 $\times$ 2$.\hspace{-2pt}'$05 
region of NGC 5033.
Gray-scale ranges from 0.5 to 1.5 mag (whiter is redder), 
and contour levels are 2, 4, 6, $\cdots$, 12 $\sigma$,
where 1$\sigma$ = 2.0 Jy beam$^{-1}$ km s$^{-1}$ or 
13 K km s$^{-1}$ in $T_{\rm b}$.
A circumnuclear ring with a diameter of $\sim$ 1.3 kpc 
and two spiral arms are clearly depicted,
and the CO distribution agrees with the ring and two spirals.
}\label{fig:colorvsCO}
\end{figure}

\begin{figure}
  \begin{center}
    \FigureFile(85mm,80mm){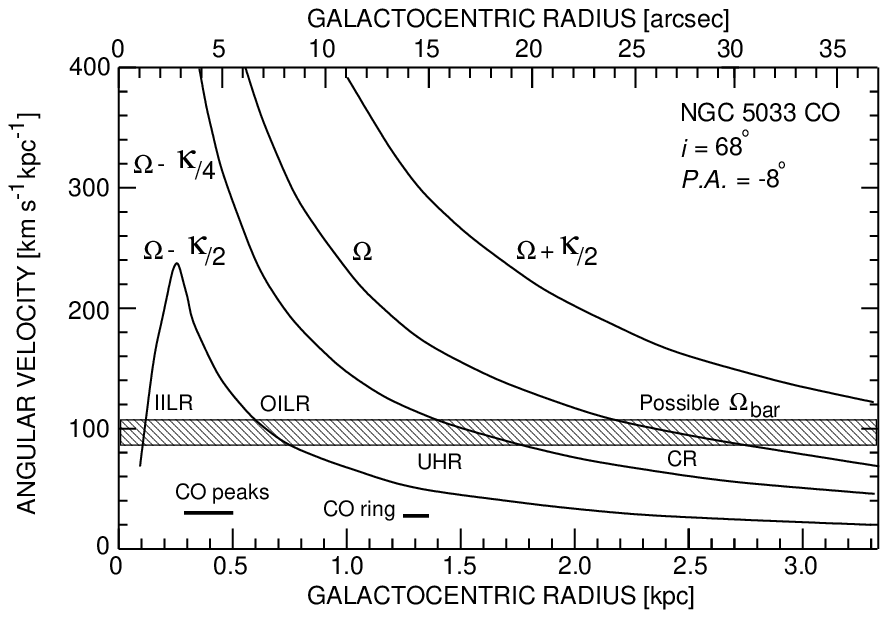}
  \end{center}
\caption{
Angular velocities ($\Omega$, $\Omega \pm \kappa/2$, and $\Omega - \kappa/4$)
as a function of galactocentric radius, 
assuming the CO circular rotation curve discussed in subsubsection 3.1.6.
A proposed range of the nuclear bar pattern speeds ($\Omega_{\rm bar}$)
is shown, which is derived by assuming the corotation resonance 
lies at the observed cut off radius of the molecular gas 
(i.e., $R_{\rm CR}$ = 2.7 kpc) 
or at the local dip radius of the CO radial distribution in figure 5
(i.e., $R_{\rm CR}$ = 2.1 kpc).
%
%
This diagram suggests that the CO ring at $r \sim$ 1.3 kpc 
corresponds to the UHR,
and the CO twin peaks and ridges occur between IILR and OILR.
}\label{fig:resonance}
\end{figure}

\begin{figure}
  \begin{center}
    \FigureFile(120mm,80mm){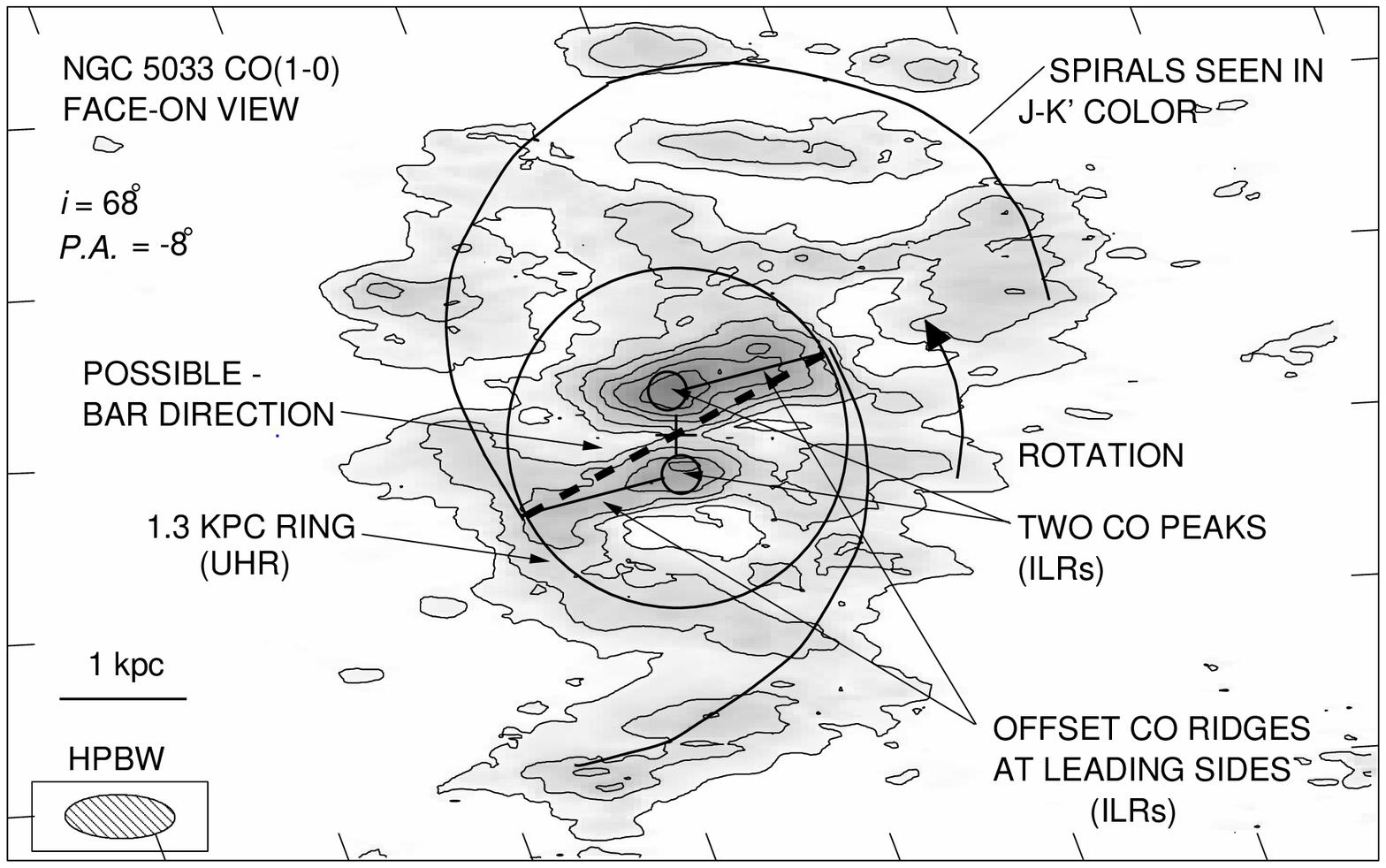}
  \end{center}
\caption{
Face-on view of the integrated CO intensity map 
in NGC 5033 (figure 3), corrected the inclination of the galactic disk. 
The abscissa is aligned to the minor axis of the galaxy, $P.A.$ = 82$^\circ$.
The thick lines trace the molecular offset-ridges, ring, and spirals structures.
These observed CO features are well understood as a response 
to the nuclear bar (indicated by the dashed line across the nucleus).
}\label{fig:faceonview}
\end{figure}

\clearpage


\begin{table*}
\begin{center}
Table~1.\hspace{4pt}Properties of NGC 5033.\\
\end{center}
\vspace{6pt}
\begin{tabular*}{\textwidth}{@{\hspace{\tabcolsep}\extracolsep{\fill}}p{12pc}llc}
\hline\hline\\[-6pt]
Parameter & Value & Reference$^*$ \\[4pt]
\hline
Morphology \dotfill                & SA(s)c                      & (1) \\
                                   & Sbc(s)I-II                  & (2) \\
Nuclear activity \dotfill          & Type 1.5 Seyfert            & (3) \\
Position of nucleus:    &  & (4) \\
\ \ \ $\alpha$ (B1950) \dotfill & \timeform{13h11m09s.171} & \\
\ \ \ $\delta$ (B1950) \dotfill & +\timeform{36D51'30".27} & \\
$D_{\rm 25}$ $\times$ $d_{\rm 25}$ \dotfill & \timeform{10'.7} $\times$ \timeform{5'.0} & (1) \\
Position angle \dotfill            & $-8^\circ$ (North = $0^\circ$, West = $- 90^\circ$)    & (5) \\
Inclination angle \dotfill         & $68^\circ$ (edge-on = $90^\circ$) & (5) \\
Systemic velocity (LSR) \dotfill   & 870 km s$^{\rm -1}$ & (5) \\
Adopted distance \dotfill          & 18.7 Mpc       & (6) \\
Linear scale \dotfill              & 90.7 pc arcsec$^{\rm -1}$   &   \\[4pt]
\hline
\end{tabular*}

\vspace{6mm}

\noindent $^*$ (1) de Vaucouleurs et al.\ 1991 (RC3); 
(2) Sandage, Tammann 1987 (RSA); (3) Ho et al.\ 1997a; (4) Ulvestad, Wilson 1989;
(5) Thean et al.\ 1997; (6) Tully, 1988.

\end{table*}


\begin{table*}
\begin{center}
Table~2.\hspace{4pt}NMA Observations of NGC 5033.\\
\end{center}
\vspace{6pt}
\begin{tabular*}{\textwidth}{@{\hspace{\tabcolsep}
\extracolsep{\fill}}p{14pc}lc}
\hline\hline\\[-6pt]
Parameter & Value \\[4pt]
\hline
Observations period \dotfill   & 1993 Dec. -- 1994 Apr., \\
                               & 1996 Jan. \\
Phase center:           & \\
\ \ \ $\alpha$ (B1950) \dotfill  & \timeform{13h11m09s.7} \\
\ \ \ $\delta$ (B1950) \dotfill  & +\timeform{36D51'27".0} \\
Visibility calibrator \dotfill & 1308+326 \\
Line \dotfill                  & CO($J$ = 1--0)      \\
Rest frequency \dotfill        & 115.271204 GHz   \\
Observed frequency \dotfill    & 114.928 GHz      \\
Array configuration \dotfill   & AB, C, and  D       \\
Projected baseline  \dotfill   & 3.8 -- 142 $k\lambda$ or 10 -- 370 m \\
Field of view   \dotfill       & 65$''$ or 5.9 kpc       \\
Band width   \dotfill          & 319.815680 MHz \\
Velocity coverage \dotfill     & 836 km s$^{\rm -1}$ \\
Velocity resolution \dotfill   & 9.78 km s$^{\rm -1}$ \\
Synthesized beam \dotfill      & \timeform{3".9} $\times$ \timeform{3".6}, $P.A.$ $6^\circ$ \\
                               & or $350 \times 330$ pc  \\
Equivalent $T_{\rm b}$ for 1 Jy beam$^{\rm -1}$ \dotfill
                      & 6.6 K    \\
r.m.s. noise in channel maps \dotfill
                      & 30 mJy beam$^{\rm -1}$  \\
                      & or 200 mK in $T_{\rm b}$ \\[4pt]
\hline
\end{tabular*}
\end{table*}


\begin{table*}
\begin{center}
Table~3.\hspace{4pt}Kinematical parameters derived from the CO velocity field.\\
\end{center}
\vspace{6pt}
\begin{tabular*}{\textwidth}{@{\hspace{\tabcolsep}
\extracolsep{\fill}}p{14pc}lc}
\hline\hline\\[-6pt]
Parameter & Value \\[4pt]
\hline
Kinematical center:     & \\
\ \ \ $\alpha$ (B1950) \dotfill & \timeform{13h11m09s.23} $\pm$ \timeform{0s.083} \\
\ \ \ $\delta$ (B1950) \dotfill & +\timeform{36D51'29".9} $\pm$ \timeform{1"} \\
Position angle \dotfill    & $-$\timeform{7D.5} $\pm$ \timeform{1D} \\
Inclination angle \dotfill & $66^\circ \pm 3^\circ$ \\
Systemic velocity \dotfill & 882 km s$^{\rm -1}$ $\pm$ 5 km s$^{\rm -1}$ \\[4pt]
\hline
\end{tabular*}
\end{table*}

\end{document}